\documentclass{ieeeoj} 

\usepackage{cite} 
\usepackage{amsmath,amssymb,amsfonts}
\usepackage{algorithmic}
\usepackage{graphicx}
\usepackage{textcomp}

\def\OJlogo{}

\def\BibTeX{{\rm B\kern-.05em{\sc i\kern-.025em b}\kern-.08em
    T\kern-.1667em\lower.7ex\hbox{E}\kern-.125emX}}

\usepackage{subcaption}
\usepackage{adjustbox}
\usepackage{booktabs}
\usepackage{multirow}
\usepackage{makecell}
\usepackage{hhline}
\usepackage{array}
\usepackage{booktabs}
\usepackage{siunitx}
\usepackage[table]{xcolor} 
\usepackage{gensymb}
\usepackage{color,soul}
\usepackage{float}
\usepackage[normalem]{ulem}
\usepackage{etoolbox}
\usepackage{indentfirst}
\usepackage{xurl}
\usepackage[hidelinks]{hyperref}
\usepackage{placeins}

\newcommand{\red}[1]{\textcolor{black}{#1}}

\usepackage{dblfloatfix}                 

\begin{document}
\receiveddate{Month XX, 2025}
\reviseddate{Month XX, 2025}
\accepteddate{Month XX, 2025}
\publisheddate{Month XX, 2025}
\currentdate{Month XX, 2025}
\doiinfo{OJAP.2026.3730332}

\title{Wideband HF Skywave Propagation: A Review with Link Modeling Corroborated by Propagation Data}

\author{ERIC WEBER\authorrefmark{1}, TED NOWAK\authorrefmark{1}, JOSEPH BERG\authorrefmark{1}, AND NADER BEHDAD\authorrefmark{1}, FELLOW, IEEE}

\affil{Department of Electrical and Computer Engineering, University of Wisconsin-Madison, Madison, WI 53706 USA}

\corresp{CORRESPONDING AUTHOR: Nader Behdad (e-mail: behdad@wisc.edu).}

\authornote{This is the author version of an article published in the \emph{IEEE Open
Journal of Antennas and Propagation} under a Creative Commons Attribution 4.0
International (CC BY 4.0) license. The published version is available at
doi:\,10.1109/OJAP.2026.3730332.}


\markboth{\textsc{IEEE Open Journal of Antennas and Propagation}}{Weber \textit{et al.}}

\begin{abstract}
The high-frequency (HF) band, ranging from 3 to 30 MHz, is widely used in a number of applications such as beyond-line-of-sight (BLOS) communications and over-the-horizon radar systems. \red{The long-range propagation characteristics of the HF band rely on the Earth's ionosphere as a refracting medium to establish BLOS wireless links.} However, traditional HF communications links often use channels with narrow bandwidths (typically 3 kHz), which limits the amount of data that can be transmitted. In this perspective article, we discuss the challenges of establishing wideband HF communications links with emphasis on the ionosphere, noise characteristics, antenna performance, and available bandwidth. Solar radiation drives ionospheric layering, where fluctuating electron densities determine usable frequencies that change with time of day, seasons, and latitude. Since background noise at HF decreases with increasing frequency, operating at the upper HF limit is advantageous. We present a noise prediction model following ITU Recommendation P.372-17 and highlight the role of high-directivity antennas such as Log-Periodic Dipole Arrays, in optimizing the signal-to-noise ratio of the link. \red{Using ray tracing techniques, we model several representative wideband HF links that employ near-vertical-incidence skywave (NVIS) or long-range skywave modes over distances ranging from approximately 390 to 5,200 km. We examine the performance of these links for bandwidths up to 1 MHz and present their day and night ionograms, with transmitter and receiver locations spanning near-equatorial to high northern latitudes. The modeled maximum usable frequencies are corroborated against crowdsourced FT8 reception data and direct ionosonde measurements over the same period for links for which propagation data was available.}

\end{abstract}


\maketitle


\section{INTRODUCTION}
\IEEEPARstart{S}{ince}\red{ the first transatlantic radio transmission established in 1901 by Guglielmo Marconi, wireless communications have become a boon in the modern age \cite{andersen_history_2017}.} Spurring advancements in radar and navigation, disaster recovery, medical imaging, and more. The continued development of radio technology has become essential in the 21st century. Due to the rapid advancements observed in the 20th century, some forms of radio technology have gained more fervor than others. Take High Frequency (HF) communications, using frequencies in the range of 3--30 MHz, for example. Despite playing a critical role in the early to mid-20th century, HF communications were soon overshadowed by satellite communications (SATCOM) which offered higher data rates, more reliable links, and required less experienced operators \cite{division_radio_2005}. Thus, HF radio became a niche communication mechanism that was primarily being used in military, disaster relief, amateur radio, and a few broadcasting applications towards the end of the 20th and the beginning of the 21st century.

In recent years, HF communications has garnered increased interest. As modern wireless infrastructure grows in scale and complexity, it also grows more vulnerable to failure and malicious threat actors. Due to the low cost of operation and ease in deployment, HF radios offer a complementary strategy to maintain communications mitigating potential disasters. Advancements in HF radio transceivers have improved link establishment making HF communications more accessible to inexperienced operators. Adaptive radio technology including automatic link establishment (ALE) and link quality analysis (LQA) relieve the operator of the burden of predicting viable ionospheric links, which tend to be unstable \cite{wang_hf_2018, adair_automatic_1989}. \red{HF skywave links have even attracted commercial interest in high-frequency financial trading, where a transatlantic skywave path can deliver limited-bandwidth market data with lower latency than submarine fiber-optic cables, since radio waves in air outpace light in glass fiber~\cite{schneider2018shortwave}.}

Successful HF transmissions are largely determined by the dynamic nature of the propagating medium, the ionosphere. \red{Radio waves traveling through the ionosphere will experience attenuation, and factors such as extrinsic noise sources, antenna gain, and choice of modulation schemes will ultimately determine the received signal’s signal-to-noise ratio (SNR).} Modern HF modems are designed to respond to these changes to ensure coherent links. A fundamental limitation of HF radio, however, is the operating bandwidth and maximum data rate. By today’s standards, the HF band would be considered inherently narrowband (3-30 MHz) and in the heyday of HF communications, it was necessary to allocate the channel bandwidth such that interference among neighboring channels was minimized. The International Telecommunications Union (ITU) designated the standard 3 kHz HF channel to support voice and low data rate transmissions up to 4.8 kbps as described in MIL-STD 188-110A \cite{pinck_medium-data-rate_nodate}. By contrast, today’s Wi-Fi 6 (practical) data rates exceed 600 Mbps \cite{rady_how_2024}.

In the past few decades, however, attempts have been made to use wideband HF waveforms such as the 24 kHz wide channels with 3 kHz sub-bands, supporting 120 kbps, as outlined in MIL-STD-188-110C Appendix D \cite{noauthor_interoperability_2011}. While the transition to 24 kHz channels represents a significant leap in HF data rates, scaling bandwidths further, to several hundred kilohertz or even 1 MHz, introduces a new set of complex physical and engineering constraints. At these wide bandwidths, the ionospheric channel is no longer flat; signals are subject to severe frequency-selective fading and dispersion as different frequency components refract from different effective heights. Furthermore, because noise power accumulates over bandwidth, maintaining a usable SNR becomes increasingly difficult without significant increases in transmit power. Finally, the physical infrastructure itself poses a hurdle, as antennas must maintain stable impedance and radiation patterns supporting the required takeoff angles (TOA) across multi-octave spans. In this perspective article, we review these issues and present our opinions on the fundamental challenges and opportunities for realizing next-generation wideband HF communications.

\section{HF COMMUNICATION LINKS}
\label{sec:links}

\subsection{Ionosphere}
\begin{figure}[h]
    \centering
    \includegraphics[width=1\columnwidth]{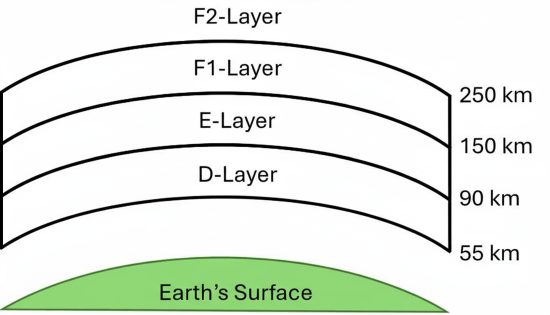}
    \caption{Layers within the ionosphere with their approximate corresponding heights. This image only describes the daytime ionosphere. During the nighttime the D- and E-layers will dissipate and the F-layer will become one large continuous layer.}
    \label{fig:IonosphereLayers}
\end{figure}

The propagation of HF radio waves is facilitated by the ionosphere, which is composed of several layers with varying degrees of ionization due to solar ultraviolet radiation. Ionization is the process of ejecting electrons from atoms.  Within these layers, positive ions and free electrons create a refractive medium in which electromagnetic radiation incident at an oblique angle will bend. Given a sufficient angle of incidence, these waves are bent back to the Earth’s surface. \red{The degree of bending at a specific wavelength or frequency reflects the state of solar illumination, and therefore varies with local time, season, solar activity, and the Earth--Sun distance. A more strongly ionized layer refracts a given frequency more sharply, and, within a given layer, lower frequencies are refracted more sharply than higher ones.} The ionospheric layers of interest to HF are the D-layer at $\sim$55--90~km above the Earth’s surface, the E-layer between $\sim$90--150~km, and the F-layer at $\sim$150+~km \cite{silver_arrl_2022}. A graphical representation of the ionospheric layers is shown in Fig. \ref{fig:IonosphereLayers}.

Ions within these layers slowly recombine. The rate of recombination is determined by air pressure, or altitude. Hence, recombination occurs rapidly in the D- and E-layers, which necessitates constant solar radiation to maintain the layer. In fact, the D- and E-layers are ionized only during the daytime hours, with ionization levels dropping rapidly at sunset. While the D-layer is not critical for HF links, it may cause significant attenuation to radio waves passing through to the F-layer, particularly at lower frequencies \cite{silver_arrl_2022}.

During the daytime hours, the F-layer is composed of the F1 ($\sim$150--250~km) and the F2-layer ($\sim$250+ km), with the F2-layer being the dominant facilitator of radio communications. At night, these layers merge forming one large layer. Due to the low air pressure at these altitudes, significant ionization is always present, making the F-layer critical for long range HF communications. \red{Again, the ion density of the F-layer varies throughout the day, rising through the morning to a maximum near noon and diminishing through the afternoon and evening hours.} These fluctuations in ion density will determine the range of frequencies which can be used for establishing links.

Fluctuations in ion density will ultimately dictate the frequencies at which HF radios can operate. \red{These densities rise and fall with local solar time, change with the seasons, and track the phase of the solar cycle.} To monitor usable frequencies throughout the day, an ionospheric sounder, or ionosonde, is used. An ionosonde is composed of an antenna that radiates vertically upwards, transmits short pulses, and detects reflected signals from the different layers of the atmosphere. The highest frequency that is returned is known as the critical frequency and is typically designated for the E-layer ($f_{o\mathrm{E}}$) and F-layers ($f_{o\mathrm{F1}}$, $f_{o\mathrm{F2}}$). As ion density increases for a layer, so too does the critical frequency. D-layer reflections happen at much lower frequencies and, therefore, D-layer critical frequencies may not always be reported. This information is reported in an ionogram. Ionograms are typically approximations as the speed of the wave is taken as the speed of light. This will not be true as the speed of the wave will slow in the presence of high ion density media. To more accurately model the ionosphere, specialty software is needed such as Automatic Real-Time Ionogram Scaler with True Height (ARTIST) \cite{silver_arrl_2022}.

 Frequencies above the critical frequency will be radiated into space unless transmitted at oblique angles with respect to the ionosphere. These oblique incidence waves give rise to the possibility of higher frequency communications. The limit to the highest possible frequency is called the maximum usable frequency (MUF) and is a function of the angle of incidence \cite {silver_arrl_2022}. The American Radio and Relay League (ARRL) defines the MUF as, ``the highest frequency supported by the ionosphere for reliable communications between two stations [for a given path]." Mathematically, the MUF is determined by

\begin{equation}
    MUF = \frac{f_o}{\sin{\theta}}
    \label{eqn:MUF}
\end{equation}

\noindent where $f_o$ is the critical frequency of a given layer and $\theta$ is the takeoff angle (TOA) with respect to the Earth’s surface. The TOA is the angle at which radio waves are transmitted and will vary depending on the antenna height, ground conductivity, and frequency. For long distance transmissions, lower TOAs are required, but will result in the waves traveling further through the D and E-layers, resulting in increased attenuation.

Similarly to the MUF, there is a lowest usable frequency (LUF). As will be discussed in a succeeding section, environmental noise is inversely proportional to frequency and ionospheric absorption is inversely proportional to the square of frequency \cite{silver_arrl_2022}. Because of this, there exists a minimum frequency where the SNR ratio becomes too low for coherent reception. The ARRL defines the LUF as, ``the frequency nearest the point where reception become[s] unusable." Lowering the LUF creates a larger usable frequency range between the MUF and LUF. The LUF can be lowered by increasing transmission power, or directivity of the antenna, thereby increasing the SNR at the receiver. In certain cases, \red{the LUF may exceed the MUF, in which case no frequency supports the link for that path.}

\subsection{Weather}
\label{sec:weather}
Both terrestrial and space weather impact HF communications, but in significantly different ways. Terrestrial weather, which occurs in the troposphere, typically plays a secondary role. While local phenomena like lightning can intermittently interrupt a link and precipitation static can add noise, their combined impact does not typically render an entire frequency band unusable. Space weather, however, has a far greater and more direct impact. By altering the ionosphere, it can close communication channels across the entire HF band, primarily by affecting the MUF.

The MUF of an HF link is governed by ionospheric conditions, which are in turn dictated by solar activity. The dominant driver of this variability is the sun, which has two cycles of primary importance. The first is the 11-year solar cycle, transitioning between solar maximum and solar minimum. Solar maximum is characterized by the presence of many sunspots—dense, cool areas of high magnetic activity \cite{solanki_sunspots_2003}—and a higher frequency of solar flares, which are massive ejections of electromagnetic energy \cite{priest_magnetic_2002}. During solar maximum, increased solar radiation, particularly Extreme Ultraviolet (EUV) radiation, energizes the ionosphere, leading to higher ion densities. This raises the MUF of the F-layers and improves global communications. Conversely, solar minima result in a weaker, less-energized ionosphere and lower MUFs \cite{australia_spaceWeather}.

Superimposed on this 11-year cycle is the sun's 27-day rotation. Because the sun is a plasma, its equator rotates faster (roughly 24 days) than its poles (up to 35 days) \cite{sun_rotation}. Since active regions like sunspots must be facing Earth to directly irradiate the ionosphere, this rotation causes additional periodic variability in propagation conditions. To quantify this solar activity for propagation models, two primary indices are used, the sunspot number (SSN) \cite{sunspot_numbers} and the 10.7 cm solar flux index (SFI) \cite{f10p7_RadioEmmissions}. The SFI, a measurement of solar radio noise at 2800 MHz, is a reliable proxy for EUV radiation. Both SFI and SSN are correlated with the 11-year cycle \cite{NOAA_SolarCycleProgression} and serve as inputs for models that predict the ionosphere's state and, consequently, the MUF for a given path \cite{wang_MUFcomparison_2024}.

While high solar activity generally increases MUFs, it is also the source of major propagation disturbances. When large solar flares occur, significant X-Ray radiation is emitted. Traveling at the speed of light, this radiation arrives with no warning, ionizing the D-layer of the ionosphere on the daylight side of Earth. This creates a ``Sudden Ionospheric Disturbance" (SID) that strongly absorbs HF signals, causing a complete signal fade-out that can last from minutes to hours, known as a radio blackout \cite{silver_arrl_2022, uryadov_impact_2018}. Solar radiation storms, also initiated by solar flares, consist of high-energy protons. These protons are guided by the Earth's magnetic field lines and funnel into the polar regions. There, the protons ionize the lower D-layer, causing a phenomenon known as Polar Cap Absorption (PCA). A PCA event can absorb HF signals traversing the polar regions for hours or even days, effectively blacking out trans-polar communication routes \cite{hakura1968polar, hargreaves_new_2005}.

Geomagnetic storms are primarily caused by Coronal Mass Ejections (CMEs), which are large-scale ejections of plasma and magnetic fields. For a storm to occur, a southward-facing Interplanetary Magnetic Field (IMF) is often required, as this allows for efficient energy transfer from the solar wind into Earth's magnetosphere \cite{lakhina_geomagnetic_2016}. This energy input heats the high-latitude atmosphere and causes it to expand. This expansion alters the neutral atmosphere's chemical composition in the F2 layer (decreasing the O/N\textsubscript{2} ratio), which accelerates the loss of ions. This process depresses F2-layer critical frequencies and thus lowers the MUF, particularly at mid and high latitudes \cite{danilov2001effects}. Additionally, geomagnetic storms expand the auroral oval, bringing strong auroral absorption effects, caused by a strong enhancement of particle precipitation, to lower latitudes than normal. Since CMEs are composed of matter, they travel slower than light, with delays of tens of minutes to several days, allowing for advanced warning.

Predicting this solar activity is extremely difficult as it fluctuates significantly day-to-day, and no two solar cycles are the same. To gain a better understanding of the sun's complex processes, scientists have launched several solar probes, such as the Parker Solar Probe \cite{raouafi_parker_2023}, the SOlar and Heliospheric Observatory (SOHO) \cite{domingo_soho_1995}, and the Solar TErrestrial RElations Observatory (STEREO) \cite{kaiser_stereo_2007}, which aim to collect data at extremely close orbits to the sun.

While space weather dictates the MUF of an HF link, terrestrial weather is the primary driver of the LUF. Terrestrial weather phenomena, occurring entirely within the troposphere, are the main contributors to this noise. Precipitation static from falling rain or snow can create significant local noise. More critically, lightning discharges from thunderstorms are a powerful source of broadband RF noise in the HF spectrum \cite{kotaki_global_1984}. This atmospheric noise can propagate thousands of kilometers through the ionosphere, particularly in summer months and in equatorial regions, raising the noise floor for receivers far from the storm itself \cite{silver_arrl_2022}. Thus, while terrestrial weather rarely ``closes" the ionosphere like space weather, it can effectively render a channel unusable by making the signal unreadable. 

\subsection{Noise}
\label{sec:noise}
As every Ham radio enthusiast knows, HF noise decreases as a function of frequency \cite{leferink_man-made_2012}. For the development of a Ham radio system, it is crucial to consider the noise power levels at the receiving antenna as we want to maximize the SNR. Obviously, one could considerably boost the input power to the transmitting antenna, but this would lead to the need for specialized hardware, including high power amplifiers and antennas that are able to withstand the desired power levels. \red{A noise power calculation factored into a link budget analysis is crucial for the development of a reliable HF radio.} One can expect higher noise levels as the bandwidth of the channel is increased \cite{sorecau_man-made_2022}.

There are three important categories of noise in the HF band: man-made, galactic and atmospheric. Man-made noise is a product of unintended electrical radiation \cite{leferink_man-made_2012, noauthor_recommendation_nodate-1}. These emitters can include automotive emission systems, electrical machinery, power lines, and combustion engines \cite{leferink_man-made_2012, noauthor_recommendation_nodate-1, noauthor_background_2017}. Each source can be further categorized into either white Gaussian noise or impulsive noise. Impulsive noise sources are ignition circuits and switching elements causing sparks \cite{leferink_man-made_2012}. Man-made noise has fluctuated throughout recent history as levels were lowered by buried power lines and newer automotive ignition systems, but the advent of personal computers and renewable power generation stations show a potential increase \cite{leferink_man-made_2012, noauthor_background_2017}. The highest man-made noise levels are found within city centers, and lowest in quiet, rural environments \cite{leferink_man-made_2012, noauthor_recommendation_nodate-1}. The difference in local environment can be taken into account within a noise model.

The second noise source is galactic noise. Galactic noise is generated by our sun and other stars in the Milky Way galaxy \cite{hoffmeyer_wideband_nodate, noauthor_recommendation_nodate-1}. The contribution of galactic noise is mostly present at higher frequencies as the lower frequency noise is attenuated by the ionosphere \cite{hoffmeyer_wideband_nodate}. Because of the degradation of the ionosphere during the nighttime, galactic noise is typically higher during this time.

The final category of noise sources is atmospheric noise. Atmospheric noise is a product of lightning strikes and thunderstorms \cite{hoffmeyer_wideband_nodate, noauthor_recommendation_nodate-1, noauthor_background_2017}. Most of the thunderstorm activity is present in three geolocations, South and Central America, Africa and Indonesia \cite{hoffmeyer_wideband_nodate}. Local lightning strikes propagate through ground waves and distant thunderstorms propagate through the ionosphere \cite{hoffmeyer_wideband_nodate}. The atmospheric noise level experiences diurnal, seasonal, and directional variations. The daily variations are a product of the degraded nighttime ionospheric density as well as increased thunderstorm activity, leading to more noise sources with less attenuation by the D-layer of the ionosphere \cite{hoffmeyer_wideband_nodate}.

The noise model presented here is from the ITU-R P.372-17 \cite{noauthor_recommendation_nodate-1}. \red{Models for man-made, atmospheric and galactic noise are each combined to provide a total estimated noise figure at the receiving antenna.} The man-made noise model is presented as median values for a short, vertical, lossless, grounded, monopole antenna. It should be noted that the man-made noise figure only considers the Gaussian distribution component of man-made noise. The impulsive component is not considered. Because impulsive noise is not considered, noise levels can become significantly higher than the level indicated by the model for a short period \cite{leferink_man-made_2012, sorecau_man-made_2022}. The median man-made noise figure follows the equation,

\begin{equation}
    F_{am} = c-d\log(f) \quad [\text{dB}]
\end{equation}

\red{\noindent where \emph{f} is the frequency expressed in MHz and $\log()$ is a base 10 logarithm}. The values of \emph{c} and \emph{d} can be found in Table \ref{tab:noise_env} and are dependent upon the environment of the receiving antenna. The provided environments are city, residential, rural and quiet rural. Upper ($D_\mathrm{u}$) and lower ($D_\mathrm{l}$ ) deviations, in units of dB, of the noise figure are also provided. These values are used when combining all noise sources together. This model is valid from 0.3 to 250 MHz, much broader than the 3--30 MHz we are considering for HF.

\begin{table}[htbp]
  \centering
  \caption{Constants for man-made noise dependent on environment, from ITU-R P.372.}
    \begin{tabular}{p{5.545em}rrrr}
    \toprule
    Environment & \multicolumn{1}{c}{$c$} & \multicolumn{1}{c}{$d$} & \multicolumn{1}{c}{$D_\mathrm{u}$} & \multicolumn{1}{c}{$D_\mathrm{l}$} \\
    \midrule
    City  & 76.8  & 27.7  & 11.0  & 6.7 \\
    Residential & 72.5 & 27.7 & 10.6 & 5.3 \\
    Rural & 67.2  & 27.7  & 9.2   & 4.6 \\
    Quiet Rural & 53.6 & 28.6 & 9.2$^{\mathrm{a}}$ & 4.6$^{\mathrm{a}}$ \\
    \bottomrule
    \end{tabular}%
  \par\smallskip
  \parbox{\columnwidth}{\footnotesize $^{\mathrm{a}}$Decile deviations for the quiet-rural category are not tabulated in ITU-R P.372; here they are assumed equal to the rural values.}
  \label{tab:noise_env}%
\end{table}%

The median galactic noise figure is represented by~(\ref{eq:fam2}) where the upper and lower noise figure deviations are each 2 dB.  This model is valid up to 100 MHz. Galactic noise is not influential below $f_{o\mathrm{F2}}$ as it is absorbed by the ionosphere and does not reach the value provided by (\ref{eq:fam2}) until the given frequency is about three times higher than $f_{o\mathrm{F2}}$. To properly model the galactic noise, a piecewise function must be created, where the noise figure below the $f_{o\mathrm{F2}}$ frequency is zero. The noise figure then would slope linearly from zero at $f_{o\mathrm{F2}}$ up to the noise figure value with the frequency set to three times $f_{o\mathrm{F2}}$. The final segment would be the resultant value from~(\ref{eq:fam2}) with the corresponding frequency.

\begin{equation}
    F_{am} = 52-23\log(f) \quad [\text{dB}]
    \label{eq:fam2}
\end{equation}

The final noise component considered within this model is the atmospheric noise. This model does not have a formula model, but rather is found by referencing world chart values of the atmospheric noise provided in ITU-R P.372-1 \cite{noauthor_recommendation_nodate-1}. In the ITU-R P.372-17 Figs. 13 – 36 give expected atmospheric noise figure values in four-hour increments in winter, spring, summer and fall \cite{noauthor_recommendation_nodate-1}. Each graph has multiple curves representing confidence intervals for the noise figure. Here each curve represents the percentage of time the noise is expected to exceed the curve (i.e., the atmospheric noise would exceed the 50\% curve half of the time). The deviation values can be similarly found in these curves although they do not contain confidence intervals but rather a single curve which gives values versus frequency.

With the model estimation of man-made, galactic and atmospheric noise completed, we can then combine each noise source to determine an overall noise figure. Each noise source must be taken into account as they generally have similar magnitude across the HF band. ITU-R P.372-17 provides formulas for the combination \cite{noauthor_recommendation_nodate-1}. Each of the noise sources is assumed to have a half-normal distribution on each side of the calculated median value $F_{am}$. The corresponding standard deviation is the provided deviation value divided by 1.282. The combined noise figure (in dB) is

\begin{equation}
    F_{amT} = 4.343 \left[ \ln(\alpha_T) - \frac{\sigma_{T}^{2}}{2(4.343)^2} \right] \quad [\text{dB}].
\end{equation}

The combined standard deviation is

\begin{equation}
    \sigma_T = 4.343 \sqrt{\ln \left( 1+\frac{\beta_T}{\alpha_{T}^{2}} \right)} \quad [\text{dB}]
\end{equation}

\noindent where

\begin{equation}
    \begin{split}
        \alpha_T &= \sum_{i=1}^{n} \exp\!\left[ \frac{F_{ami}}{4.343} + \frac{\sigma_i^2}{2(4.343)^2} \right] \quad [\text{W}] \\
        \beta_T &= \sum_{i=1}^{n} \alpha_i^2 \left[ \exp\!\left( \frac{\sigma_i^2}{(4.343)^2} \right) - 1 \right] \quad [\text{W}^2].
    \end{split}
\end{equation}

$F_{ami}$ and $\sigma_i$ are the median noise figure and standard deviation values for each noise source respectively. In the case of an individual deviation exceeding 12 dB the combined standard deviation should be restricted to

\begin{equation}
    \sigma_T = 4.343\sqrt{2 \ln\left(\frac{\alpha_T}{\gamma_T}\right)} \quad [\text{dB}]
\end{equation}

\noindent where

\begin{equation}
    \gamma_T = \sum_{i=1}^{n} \exp\!\left[ \frac{F_{ami}}{4.343} \right] \quad [\text{W}].
\end{equation}

\noindent Once the combined noise figure has been determined it is then used to calculate the noise power. This noise power value can be used in a link budget analysis to determine the SNR. The noise power is calculated by

\begin{equation}
    P_n = F_{amT} + 10 \log(BW) + 10\log(kT_0) \quad [\text{dBW}]
    \label{eqn:noise_power}
\end{equation}

\noindent where $BW$ is the bandwidth in Hz, \(k\) is Boltzmann’s constant ($1.38\times10^{-23}$   [J/K]), and $T_0$ is the reference temperature, taken to be 290~K.

\begin{figure}[!t]
    \centering
    \includegraphics[width=1\columnwidth]{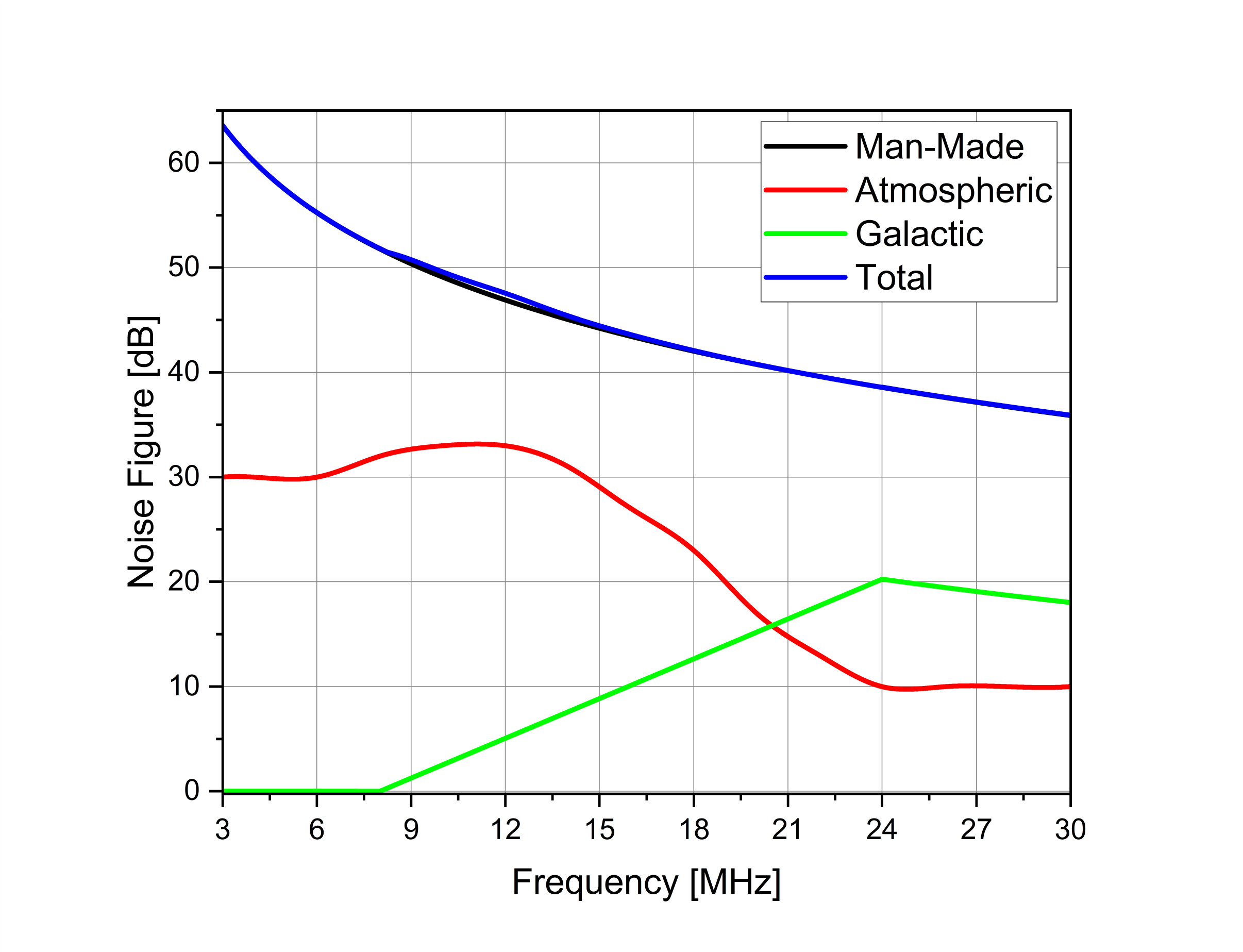}
    \caption{Man-made, atmospheric, galactic, and total noise figures for a northern link during summer conditions. The $f_{o\mathrm{F2}}$ value was set to 8 MHz, for a typical summer day, with the environment defined as a city-center. The total noise closely follows the man-made component due to the elevated noise levels characteristic of city environments.}
    \label{fig:NorCitSum}
\end{figure}

\begin{figure}[!t]
    \centering
    \includegraphics[width=1\columnwidth]{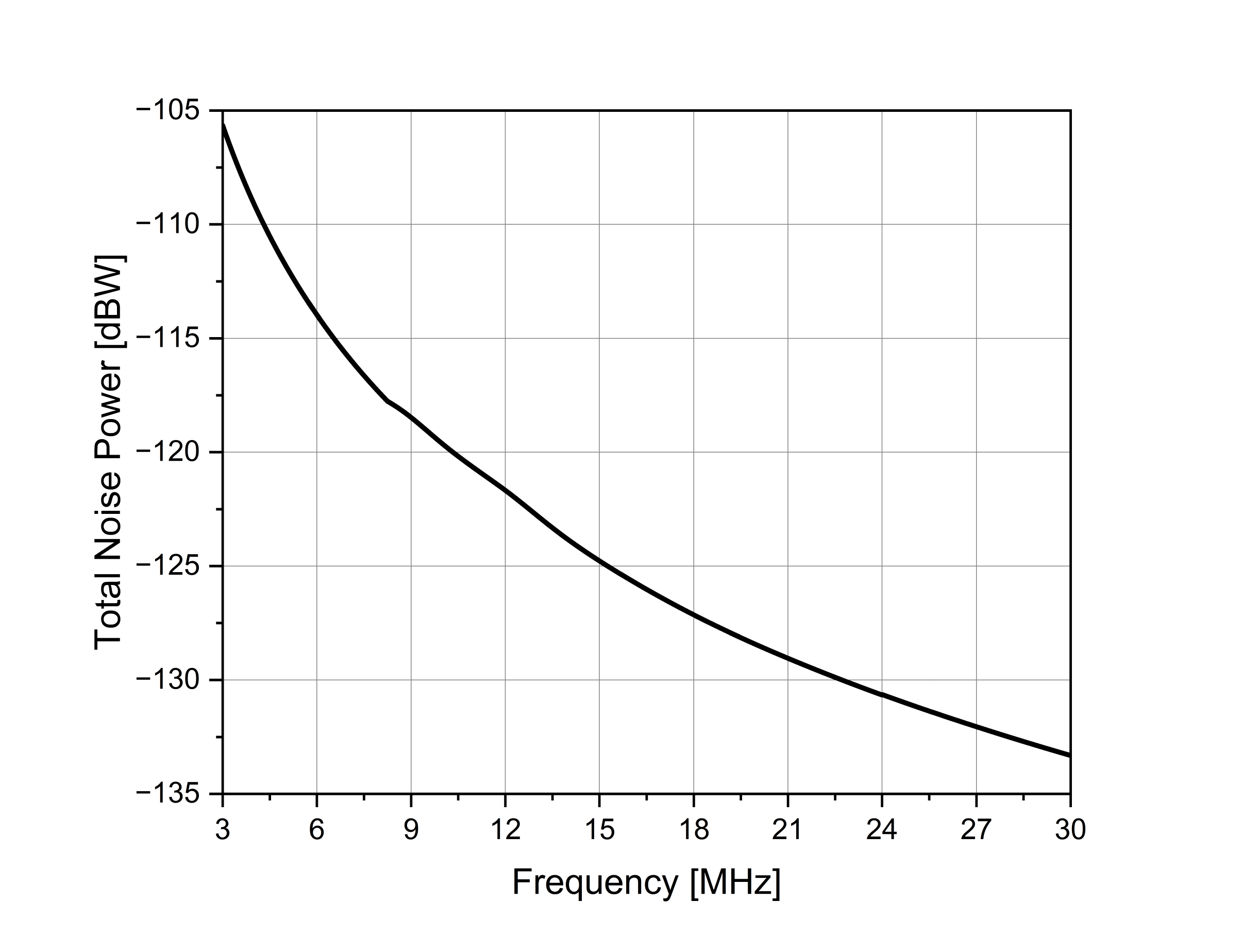}
    \caption{Corresponding total noise power to Fig. \ref{fig:NorCitSum} with a 3 kHz bandwidth.}
    \label{fig:NorCitSumPower}
\end{figure}

An example of the calculated noise figure and noise power is shown in Figs. \ref{fig:NorCitSum} and \ref{fig:NorCitSumPower}, respectively. The presented results correspond to a northern link in summer, with the receiver in a city center, assuming a daytime $f_{o\mathrm{F2}}$  of 8 MHz. A 3 kHz bandwidth was selected for the noise power calculation. As shown, the total noise figure is primarily dominated by man-made noise. The man-made noise component alone can be reduced by nearly 20 dB when moving the receiver into a quiet rural environment. The noise characteristics for the northern and southern links are largely similar, with the primary difference arising from galactic noise, which decreases in southern locations due to the higher $f_{o\mathrm{F2}}$ values that elevate the effective reflection frequency and reduce galactic noise contribution.

\subsection{Link Budget Analysis}

\hspace{6pt} A link budget analysis can be performed to establish the feasibility of skywave and ground wave propagation paths. Following the calculation of path loss for skywave links and basic transmission loss for ground wave links, a simplified link budget can be calculated in a software such as MATLAB or Python. The link budget presented here considers the transmit power, antenna gains, and losses occurring strictly between the transmitter and receiver. It does not account for insertion losses due to RF components, cable losses, or return loss due to mismatches in the system. The link budget is calculated as follows:

\begin{equation}
    P_{rx} = P_{tx} + G_{tx} - L_{tx} + G_{rx} - L_{rx} - PL.
    \label{eqn:lba}
\end{equation}

\noindent Here, $P_{rx}$ is the received power in dBm, $P_{tx}$ is the transmit power in dBm, $G_{tx}$ and $G_{rx}$ are the antenna gains in dBi, $L_{tx}$ and $L_{rx}$ are additional system losses, and $PL$ is the path loss. Additional system losses include the losses of cables and connectors, mismatch losses between different components, and polarization loss at the receiving antenna. Path loss can be obtained from HF propagation path simulations for skywave links using software packages such as PropLab or PHaRLAP, and basic transmission loss can be obtained from the LFMF model for ground waves \cite{PropLabPro_V3.2, DSTG_PHaRLAP, NTIA_LFMF}. After calculating the received signal power and the noise power using the model described in the preceding section, the receiver’s SNR is calculated using

\begin{equation}
    \mathrm{SNR}~[\mathrm{dB}] = P_{rx}~[\mathrm{dBm}] - P_n~[\mathrm{dBm}].
\end{equation} 

\section{HF ANTENNA SELECTION}
\label{sec:antennas}
\indent HF communications presents a mixture of opportunities and challenges for antenna design, selection, and ease of use. On one hand relatively simple wire antennas such as monopoles, dipoles, loops, and inverted-V configurations can be used to establish skywave and groundwave links. On the other hand, the long wavelengths of electromagnetic waves at the HF band impose substantial physical constraints on the antenna size. At the lower end of the band, a resonant half-wave dipole may approach 50 meters in length and vertical monopoles require extensive ground systems to obtain adequate performance \cite{silverARRLAntennaBook2019}. \red{Additionally, the presence of earth in the vicinity of the antenna can significantly impact its performance metrics that are important in wideband HF communication systems.} These include gain, bandwidth, efficiency, and the shape of its radiation pattern determining gain values at the takeoff angles required to establish a given communication link. Additionally, site selection, infrastructure, and mechanical support often become major design considerations.

In wideband HF communications, a further complication arises from the inherent narrow bandwidth of resonant wire antennas. While such antennas are efficient at their design frequency, their impedance, radiation pattern, and efficiency can change significantly across the spectrum. Modern HF systems have been designed to mitigate  some of these limitations.  Antenna tuning units (ATUs) allow electrically short or frequency-selective antennas to operate acceptably across wide frequency ranges. ALE further improves usability by dynamically identifying the optimal operating frequency. Moreover, research in antenna miniaturization techniques aims to reduce the overall footprint of HF antennas  \cite{balanisAntennaTheoryAnalysis2016, zhuMiniaturizedTransmittingLPDA2021}. These advancements, however, have largely been in the context of HF communications using channel bandwidths less than 24 kHz. Despite these advancements, many common HF antenna types are not able to provide significantly wide instantaneous bandwidths that may be necessary if radiation of  waveforms with bandwidths in the range of several hundred kHz to around 1 MHz are desired. In contrast, broadband antenna designs such as terminated folded dipoles, LPDAs, and traveling-wave structures provide relatively wide operating bandwidth and stable radiation patterns \cite{silverARRLAntennaBook2019, balanisAntennaTheoryAnalysis2016, carrPracticalAntennaHandbook2001, stutzmanAntennaTheoryDesign2013}, which makes them attractive options for wideband HF communications. However, these attractive features come at the expense of reduced efficiency for terminated folded dipoles \cite{silverARRLHandbook2023_ch21} or relatively large physical dimensions in the case of LPDAs and traveling wave antennas such as long wire or beverage antennas \cite{stutzmanAntennaTheoryDesign2013, silverARRLAntennaBook2019_ch7, LongWireAntennasPart, Carr1998Beverage}.  \red{At the lower end of the HF band an efficient antenna is unavoidably electrically small, and the bandwidth--efficiency product of any passive, \emph{linear time-invariant} (LTI) electrically small antenna is bounded by well-established physical limits, making efficient radiation of the several-hundred-kHz to 1-MHz waveforms of interest here fundamentally difficult. A promising route beyond this bound is to relax the linear time-invariance assumption on which the limit rests by using non-LTI (i.e., time-varying and/or nonlinear) electrically small antennas. Recent work has demonstrated such an approach, integrating an electrically small antenna directly with a switched-mode power amplifier, with no intermediate $50~\Omega$ impedance-matching network, and dynamically modulating the antenna quality factor to yield a compact HF transmitter with an enhanced bandwidth--efficiency product. Using this technique, over-the-air transmission of wideband amplitude-shift-keyed (ASK) and phase-shift-keyed (PSK) waveforms, including binary (BPSK), quadrature (QPSK), and 8-ary (8PSK) variants, at symbol rates up to 1~MHz has been achieved from a $\sim$1-m electrically small HF monopole near 10.7~MHz~\cite{ma2025overair, ludois2026qmod}.}

\subsection{Site Selection}

Proper site selection is imperative, as it strongly influences feasible antenna types and achievable link performance. Site planning, as described in ITU Rec. F-1610, involves choosing a location that is low-noise, flat, and relatively unobstructed, with suitable ground conductivity \cite{ITU-R-F.1610-2003}. The site must have basic infrastructure, including access to power and sufficient land or clearance for the intended antenna systems, while towers or ground radials may be required depending on the antenna type. Ultimately, the goal in site selection is to maximize the signal-to-noise ratio and ensure practical means of antenna installation and maintenance.

\subsection{Antenna Considerations}

Antenna selection will begin with identifying the desired operating frequencies and the required bandwidth. These choices impose several constraints, namely the type of antenna that can be used (narrowband resonant, multiband, or broadband) and the physical size of the antenna. Narrowband antennas include familiar resonant structures such as the half-wave dipole, quarter-wave monopole, and Yagi–Uda \cite{silverARRLHandbook2023_ch21, silverARRLAntennaBook2019_ch11}. Multiband options include trapped dipoles or verticals, fan dipoles, and end-fed half-wave antennas that take advantage of harmonic resonances \cite{silverARRLAntennaBook2019_ch10}. Broadband HF antennas include the log-periodic dipole array and terminated designs such as the tilted terminated folded dipole \cite{silverARRLHandbook2023_ch21, palomarBBTDproducts, silverARRLAntennaBook2019_ch10}.

The choice of antenna is further narrowed by determining the desired link. The expected coverage, whether it be point-to-point, long-range skywave, or NVIS, will dictate the required radiation characteristics such as polarization, takeoff angle, and directivity \cite{silverARRLAntennaBook2019, lapinARRLHandbookRadio2024}. For example, ground wave links necessitate vertically polarized antennas, whereas long-range skywave is effectively agnostic to polarization but requires low TOA \cite{dolukhanov_propagation_1995, itziar_angulo_handbook_nodate}. It is also well known that directive antennas improve long-distance reception by increasing gain and rejecting interference. Conversely, for local communications or disaster-relief scenarios, omnidirectional antennas are better suited to provide wide-area coverage \cite{ITU_R_BS_80_3_1990, NTIA_Interference_Resilient_2025}.

Final considerations for antenna selection include determining a link budget, antenna feed and matching requirements, power handling, and the type of modulation that will be used. A link-budget analysis can estimate the antenna gain and transmit power needed for a desired SNR, though the performance of any link can significantly change as a result of ionospheric variability and time of day. Depending on the antenna architecture, a proper feed and/or matching network (or ATU) may be required. For example, a half-wave dipole is a balanced structure and is typically fed through a balanced-to-unbalanced (balun) transformer when using coaxial feed lines. A balun can also transform between the feed-line impedance and the antenna impedance to maintain an acceptable voltage standing-wave ratio (VSWR) \cite{balanisAntennaTheoryAnalysis2016}. While feed-line losses at HF are generally modest, low-loss lines are still recommended. Although most HF communications are externally noise-limited, excessive losses in long feed lines can further reduce already weak signals and degrade performance on low-SNR paths \cite{noauthor_recommendation_nodate-1}. An ATU may be added to extend usable bandwidth or accommodate frequency changes. Finally, the antenna must be rated to handle the transmit power of the transceiver and any external amplifiers, as well as the peak-to-average power ratio (PAPR) associated with the chosen mode.

\subsection{Wideband HF Antennas}
\label{sec:wbantennas}

Successful wideband transmissions require antennas with stable impedance and radiation characteristics over a large frequency range. This is particularly challenging at HF due to the large difference in wavelength between the lower (100 m at 3 MHz) and upper (10 m at 30 MHz) portions of the band. Unlike narrowband or multiband resonant antennas that provide acceptable performance only at specific frequencies, wideband antennas are designed to operate across multiple octaves and, in some cases, across the entire HF spectrum. Although current wideband HF waveforms occupy channels on the order of 12–24 kHz, frequency-agile systems and robust HF links still benefit from antennas capable of operating over many megahertz of the spectrum. Additionally, deployment of wideband HF systems employing bandwidths exceeding a few hundred kHz will require using wideband antennas. Multi-octave coverage allows wideband transmissions to remain effective as operating frequencies shift in response to daily variability in the ionosphere.

Across this range, antennas must maintain a good impedance match, suitable gain, and appropriate radiation patterns to establish the desired link at different frequencies. A few potentially suitable candidates include the log-periodic dipole array \cite{silverARRLAntennaBook2019_ch7}, the tilted terminated folded dipole (TTFD) \cite{silverARRLHandbook2023_ch21, palomarBBTDproducts}, and wideband vertical monopoles \cite{Sabre_XHF_2025}. The LPDA belongs to the class of frequency-independent antennas and is constructed from a series of dipole elements whose lengths and spacing vary logarithmically along a central boom \cite{balanisAntennaTheoryAnalysis2016, silverARRLAntennaBook2019}. All elements are connected to a single feed line with adjacent elements driven in phase opposition. For any given frequency, only a portion of the array is active, resulting in the characteristic stable performance across multiple octaves. LPDAs can achieve bandwidths that span the entire HF spectrum while maintaining low VSWR and high radiation efficiency. However, to accommodate the lower end of the HF band, an LPDA may become prohibitively large.

The TTFD resembles a traditional folded dipole but includes a resistor termination opposite the feed point. The termination acts as a broadband resistive damper, suppressing standing waves along the structure and effectively reducing the antenna's quality factor \cite{silverARRLAntennaBook2019}. This allows the TTFD to present a relatively stable impedance and VSWR across the band while having a relatively small size compared to LPDAs. It is not uncommon to achieve a VSWR less than 3:1 over the whole band. TTFDs are generally less efficient than resonant antennas due to power being dissipated as heat in the resistive termination. What the TTFD sacrifices in efficiency it gains in simplicity, predictable broadband response, and ease of installation.

Beyond the LPDA and TTFD, other broadband antenna designs such as the terminated long-wire \cite{LongWireAntennasPart}, rhombic \cite{HFRARhombic}, and diamond \cite{Sabre_XHF_2025} antennas can be used for wideband HF operation. Terminated long-wire antennas provide predictable broadband performance and inherently low TOAs, but their traveling-wave nature produces multiple sidelobes that may cause unwanted radiation toward undesired directions \cite{balanisAntennaTheoryAnalysis2016, LongWireAntennasPart}. Rhombic antennas offer even stronger directionality and stable wideband behavior, but their gain and TOA vary significantly across the HF spectrum and they require large physical footprints \cite{HFRARhombic}. Vertical diamond antennas may also be employed, which offer wide instantaneous bandwidths across the entire HF band, though they typically exhibit low efficiency and modest gain \cite{Sabre_XHF_2025}.

In the context of wideband HF, antenna selection becomes a central constraint on overall system feasibility. While traditional resonant designs offer very good performance for narrowband operations, they cannot provide the multi-octave coverage required for agile, wideband operation. Broadband antennas such as the LPDA and TTFD mitigate this limitation by offering stable impedance and radiation characteristics across the HF spectrum, albeit with trade-offs in size or efficiency. These considerations highlight that successful wideband HF ultimately depends on choosing antennas capable of supporting wide instantaneous bandwidths and operating over a very wide frequency band while balancing practical deployment constraints. 

\section{HF CHANNEL BANDWIDTH}
\label{sec:bandwidth}

\subsection{Narrowband Communication Modes}
A pure, unmodulated sinusoidal carrier wave transmits no information other than its own existence. To communicate, information must be encoded onto this carrier through modulation. Modulation methods can be analog, where a continuously varying signal (such as voice) is encoded, or digital, where the signal takes on discrete values (such as 1s and 0s). The specific method, or protocol, used to encode this information determines the signal's bandwidth (BW). The FCC defines bandwidth as ``the width of a frequency band outside of which the mean power of the transmitted signal is attenuated at least 26 dB below the mean power of the transmitted signal within the band" \cite{FCC97_3_a_8}. To overcome noise and transmit various types of information, a wide range of modes have been developed, balancing bandwidth, data speed (baud rate), and robustness.

The earliest method of HF communication was Continuous Wave (CW), which uses On-Off Keying (OOK) to transmit Morse code. In this modulation scheme, the carrier's amplitude is either ``on" or ``off," with information encoded in the duration of the pulses. With a typical bandwidth of around 150 Hz, CW is extremely narrowband \cite{silver_arrl_2022}. This concentration of energy makes transmissions highly efficient and robust against noise, allowing low-power signals to be decoded in environments where other modes fail, ensuring CW's continued use to this day.

As technology advanced, it became possible to transmit human voice. Early voice transmission encoded the vocal signal into the amplitude of a carrier wave. This process generates two identical signal components above and below the carrier frequency, known as sidebands. Double sideband (DSB) AM transmits the carrier and both sidebands, resulting in simple demodulation but poor power and spectral efficiency, occupying a bandwidth of approximately 6 kHz \cite{institute_for_telecommunication_sciences_required_1969}. A far more efficient method is single sideband (SSB), where the carrier and one of the sidebands are filtered out prior to transmission. This reduces the bandwidth to less than 3 kHz. Current FCC regulations and standards limit the transmitted bandwidth for amateur HF communications to 2.8 kHz \cite{noauthor_amateur_2023}.

While transmitting full-motion video over HF is unfeasible due to TV signals requiring several megahertz of bandwidth, transmitting static images is possible. The earliest method was radio facsimile, or Radiofax. Radiofax transmits grayscale images by encoding black and white levels into specific audio tones transmitted via SSB. In this way, weather charts and satellite images can be transmitted line-by-line and reconstructed at the receiving end. It remains in use today for maritime safety broadcasts. Building on the concept of Radiofax, Slow-Scan TV (SSTV) is sent at a faster rate and includes synchronization pulses \cite{silver_arrl_2022}. The standard transmitted bandwidth of SSTV is less than 3 kHz \cite{noauthor_image_nodate}.

Digital modes offer significant advantages over analog, allowing for complex encoding schemes, error correction, and extremely narrow bandwidths. One of the original digital data modes, Radioteletype (RTTY), uses Frequency-Shift Keying (FSK) to shift the carrier between two distinct frequencies representing binary 0 and 1. RTTY has a typical bandwidth of around 600 Hz, though this varies with the baud rate and frequency shift \cite{klapashchuk_analysis_2024}. Advancing on RTTY, Amateur Teleprinting Over Radio (AMTOR) incorporates Automatic Repeat reQuest (ARQ). This establishes a two-way link where the receiving station automatically checks data for errors and requests retransmission as needed. AMTOR has a broader bandwidth than RTTY, typically occupying approximately 800 Hz \cite{klapashchuk_analysis_2024}. PACTOR further improved upon AMTOR by integrating the robust error correction of packet radio with ARQ, significantly improving reception of weak or noisy signals. There are several generations of PACTOR, with bandwidths dependent on transmission speed. PACTOR 1 has a bandwidth of about 600 Hz at 100 baud \cite{klapashchuk_analysis_2024}, while the newer PACTOR 4 has a bandwidth of 2.4 kHz at 1,800 baud \cite{noauthor_pactor-4_nodate}.

Other digital modes utilize PSK, which encodes data by applying a phase shift to the carrier frequency. Popular amateur radio modes include PSK31, which has an exceptionally narrow bandwidth of 60 Hz \cite{noauthor_psk31_nodate}. This mode allows for efficient keyboard-to-keyboard communication at very low power. Finally, Multi-Frequency Shift Keying (MFSK) sends data using many different audio tones simultaneously, making it highly resistant to the fading and multipath distortion common on HF paths. The transmission bandwidth of MFSK is determined by the number of tones multiplied by twice the tone spacing \cite{noauthor_mfsk_nodate}.

\subsection{Wideband Experiments}
\label{sec:wideband}

\subsubsection{Long Range Skywave}
Previous wideband long range skywave HF experiments are limited. In the 1980’s a campaign by the National Telecommunications and Information Administration (NTIA) set out to measure, model, and simulate wideband HF channels \cite{hoffmeyer_wideband_nodate}. Much of the work was founded on research groups conducting wideband HF experiments beginning in the late 1960’s through the late 1980’s. The problem with developing an accurate model at the time was the lack of channel characterization, which did not begin until the mid-20th century. In addition, empirical data collected up to this point was predominately narrowband to validate the narrowband Watterson model \cite{watterson_experimental_nodate}. The wideband HF model that was described was essentially a generalization of the narrowband Watterson model. This was problematic because the Watterson model neglected dispersion. The authors themselves concluded that the wideband model was “more conjectural than factual” \cite{hoffmeyer_wideband_nodate}. However, several important works were documented in the NTIA report.

Several groups, including the Naval Research Laboratory (NRL), Mitre Corporation, and the Norwegian Defense Research Establishment experimented with wideband HF channels in excess of 100 kHz. The NRL conducted channel sounding and channel probing at 125 kHz and 1 MHz, respectively. The sounder was located at San Clemente Island, CA and consisted of a 25-element radar array. Channel sounding was conducted at power levels between 200 W and 3 kW.  Channel probing was conducted at power levels between 10 and 300 W between San Clemente Island and Point Mugu, CA, approximately 130 km. The researchers concluded that one-hop F-layer propagation can be conducted at bandwidths up to 1 MHz, provided optimal conditions. A full system description and design can be found in \cite{wagner_wideband_nodate}.

The Mitre Corporation experimented with adaptive spread spectrum wideband channels with coherent bandwidths of 1 MHz. The researchers examined one-hop F-layer links of approximately 2,000 km from Bedford, MA to Eglin Air Force Base, FL with the majority of the experiments taking place in the spring and summer. Spread spectrum signaling and real time ionograms allowed the researchers to select among intermodal paths (signals arriving from, for example, both the D- and E-layer) and isolate the one-hop F-layer signal. In addition, equalizers were incorporated in the design to minimize dispersion and fading due to intramodal multipath (multipath originating from ordinary and extra-ordinary modes). Equalization decreased dispersion from 95 $\mu$s/MHz to 1 $\mu$s/MHz. The researchers determined that equalized wideband channels are time invariant over a period of 10 seconds. These tests were conducted using a broadband log periodic array (LPA) with a transmit power of 500 W \cite{dhar_equalized_1982}.

Other works include those performed by Skaug and the Norwegian Defense Research Establishment. Skaug experimented with a range of bandwidths at HF including 2.5, 10, 40, 80, and 160 kHz. Experiments were conducted over an 840 km path across Norway in the autumn and winter months at a frequency of 8 MHz. The researchers were limited by the low MUF due to the higher northern latitudes. It was reported that significant drops in performance were seen at 160 kHz bandwidths. It was speculated that this drop in performance was due to the large variation in dispersion among the outermost components of the spectrum. The author concluded that bandwidths of 80 kHz and below remained coherent and acknowledged that further performance improvements could be made for higher bandwidths \cite{skaug_experiment_1984}. A detailed list of wide bandwidth experiments can be found in the NTIA report \cite{hoffmeyer_wideband_nodate}. 

\subsubsection{Near Vertical Incidence Skywave}
At the time of writing, the largest bandwidth seen employed in NVIS was 200 kHz  \cite{witvliet_radio_2017}. Many publications have shown the efficacy of the 24 kHz bandwidth and these links were implemented in both stationary and mobile applications \cite{allen_mid-latitude_nodate, ignatenko_wide-band_2016}.  However, results indicate the difficulty of implementing communications links with wider bandwidths for NVIS. For example, Allen, et al. \cite{allen_mid-latitude_nodate} discussed an NVIS channel model which indicated that receive SNR dropped from over 10 dB to less than 5 dB at 12 kHz and 25 kHz bandwidths, respectively. The models assumed a loop antenna and noise power was calculated using an urban model as a function of bandwidth. In addition, for the mobile application in question, it was determined that at 50 kHz, there would be insufficient power to achieve adequate SNR \cite{allen_mid-latitude_nodate}.

However alternate waveforms show promise for improving signal integrity. Laraway et al. \cite{laraway_experimental_2016} utilized a filter bank multicarrier spread spectrum waveform across 24 kHz and 200 kHz bandwidths. Experiments demonstrated that signals with SNR values less than -20 dB were used to establish communications. A link was established between Idaho and Utah in the spring of 2016. A USRP software defined radio (SDR) was used to create the desired waveforms. Signals were amplified between 0.5 W and 25 W and fed to a folded dipole centered at approximately 6.4 MHz \cite{laraway_experimental_2016}. The authors reported that at the 200 kHz bandwidth, signals received at approximately -16 dB were generally error free. While much of HF communications utilizes a continuous bandwidth, this need not be the case. Lamy-Bergot et al. employed a non-contiguous 200 kHz bandwidth divided into 3 kHz sub-bands. Due to the rapidly changing nature of the ionospheric channels, this offered the experimenters the ability to select the most appropriate sub-band providing the highest SNR. The researchers experimented with an approximate 350 km link using commercial off the shelf antennas. Results showed that using a larger non-contiguous bandwidth, channel availability improved by over 50\%. This technique provides a more robust and adaptive modem to the channel characteristics \cite{lamy-bergot_-air_nodate}.

\subsubsection{Ground Wave}
To the author’s knowledge, no publications exist regarding wideband high frequency ground wave propagation beyond the standard 3 kHz bandwidth. Although surface waves may experience slow fading due to irregular terrain or a nonhomogeneous surface, guided waves experience minimal dispersion at wide bandwidths \cite{fabrizio_high_2013}. Therefore, in theory, wideband surface wave propagation is possible.

\section{TYPES OF LINKS}
\label{sec:TypesOfLinks}
HF radio waves can propagate in three distinct modes: long range skywave, ground waves, and near vertical incidence skywave (NVIS). Long range skywaves utilize the ionosphere to refract radio signals back to the Earth and support communication distances in excess of 4,000 km. Ground waves are a special type of propagation mode in which radio waves travel across (half in and half out of) the Earth’s surface and bend with the curvature of the Earth. Ground waves support communication distances up to 500 km in favorable conditions \cite{sevgi_ground_2003}. NVIS is a specific type of skywave propagation characterized by large TOA’s between 40\degree  and 90\degree \cite{silver_arrl_2022}. NVIS is used to cover what is known as the \textit{skip zone}. The skip zone is the region between the longest path distance supported by ground wave propagation and the shortest path distance supported by long range skywave propagation. 

\subsection{Long Range Skywave}
Long range skywaves are by far the most important propagation mode for HF communications, but also tend to be the most unpredictable. Long range skywaves support communication distances in excess of 2,000 km depending on the layer from which signals are refracted. Single hop E-layer, F1-layer, and F2-layer support communication distances of roughly 2,000 km, 3,000 km, and 4,000 km, respectively \cite{silver_arrl_2022, australian_government_space_nodate}. Multi-hop links from the F-layer are capable of traversing Earth's circumference (about 40,000 km). The typical range of frequencies supported by long range skywaves refracted from the F-layer are 10 to 50 MHz \cite{silver_arrl_2022, noauthor_module_nodate}. Frequencies below 10 MHz tend to be absorbed by the D- and E-layers. Frequencies above 50 MHz pass through the ionosphere into space unless under very special circumstances \cite{silver_arrl_2022}.

The range of frequencies supported by the F-layer will change depending on the time of day, season, position within the solar cycle, and geography. Higher MUFs will generally be seen near the equator where higher electron densities are present. Care should be taken to avoid significant D- and E-layer attenuation, which is more prominent at lower frequencies. Lower frequencies are also more susceptible to radio noise as discussed in the preceding sections.

In general, for HF propagation, the choice of antenna and the antenna radiation pattern will also determine whether or not a given link is accessible. For long range skywaves, which require low TOA’s for long distance communications, this is especially important. The choice of antenna, polarization, frequency, the antenna height off the ground, and ground type (dry/wet soil, sea/fresh water) will all affect the antenna’s radiation pattern. In addition, for wide bandwidth links, the antenna must have a suitable match across the desired bandwidth, as described in Section \ref{sec:antennas}-\ref{sec:wbantennas}.

\subsection{Near Vertical Incidence Skywave}
NVIS is a special case of skywave propagation that allows us to bridge the \textit{skip zone}. In other words, NVIS allows for intermediate communication distances that cannot be reached by ground waves or are “skipped” by long range skywaves. NVIS is a popular choice for disaster relief or emergency situations where local communications are required. Due to the high TOA’s, NVIS is also immune to obstacles that might otherwise block transmissions such as buildings and mountains \cite{witvliet_radio_2017}. While NVIS waves can refract from both the E- and F-layers, F-layer propagation is preferred as it supports higher operating frequencies, which are less susceptible to noise and fading. With these large TOA’s, the maximum supported frequency for NVIS approaches the critical frequency of a given layer. Practically speaking, this means that the usable frequencies are much lower, typically between 2 and 10 MHz. The usable frequencies will depend on the time of day, season, and latitude.

Fundamentally, long range skywave and NVIS propagation both rely on refraction of an HF wave from the ionosphere. However, special considerations must be made for successful NVIS transmissions. For example, antennas that are typically used for long range skywaves radiate at low angles toward the horizon. For NVIS, the antenna's maximum radiation should be directed towards higher elevation angles or near vertical. TOAs in excess of 70$^\circ$ are common \cite{allen_mid-latitude_nodate}. Antennas that work well for NVIS include Rhombic antennas \cite{HFRARhombic}, log periodic conical spirals \cite{witvliet_radio_2017}, LPDA’s oriented vertically, and the inverted-V \cite{silverARRLHandbook2023_ch21}. Half-wave dipoles and loop antennas are also popular for their low price to performance ratio and are used widely for NVIS operations.

Standard channel bandwidths are 3 kHz with wideband channels up to 24 kHz being supported (MIL-STD 188-110C, Appendix D) in modern modems. The same interoperability standards (MIL-STD 188-110A) apply to NVIS. Due to the lower operating frequencies, wider bandwidths are difficult to realize with NVIS. Furthermore, noise sources such as man-made noise are heavily concentrated in the 2 to 5 MHz range, making the lower end of the NVIS frequency range sub-optimal for wideband communications \cite{lossmann_hf_2011}.

\subsection{Ground Wave}
\begin{figure*}[h]
    \centering

    \includegraphics[width=0.75\textwidth]{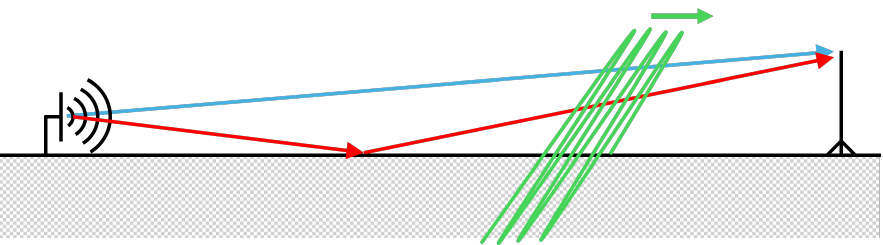}

    \caption{The complete ground wave consisting of the direct wave (blue), reflected wave (red), and surface wave (green).}
    \label{fig:gwdiagram}
\end{figure*}

For short distance communications, we have the ground wave, which is best suited for lower frequencies in the HF band (2 – 4 MHz). In the literature, the ground wave refers to three independent components: the direct wave, the reflected wave, and the surface wave as shown in Fig. \ref{fig:gwdiagram}. As the name suggests, the direct wave is a direct line of sight (LOS) link from the transmitter to the receiver. The reflected wave is incident on the receiver after reflecting off the surface of the ground and experiences attenuation and a phase shift \cite{dolukhanov_propagation_1995}. As a result, the reflected wave will interfere either constructively or destructively with the direct wave. Finally, we have the surface wave, which dominates when the antennas are positioned close to the ground. The surface wave travels along the Earth’s ground-air interface and suffers from additional attenuation due to induced surface currents. \red{In addition, the surface wave exhibits a normal and a transverse electric field component which result in a power flux density directed forward and down in the direction of propagation.} Thus, the surface wave is an elliptically polarized wave with a forward tilt that enables it to diffract and follow the curvature of the Earth, extending beyond the horizon \cite{dolukhanov_propagation_1995, itziar_angulo_handbook_nodate, fabrizio_high_2013}.

\begin{figure}[h]
    \centering

    \includegraphics[width=0.5\textwidth]{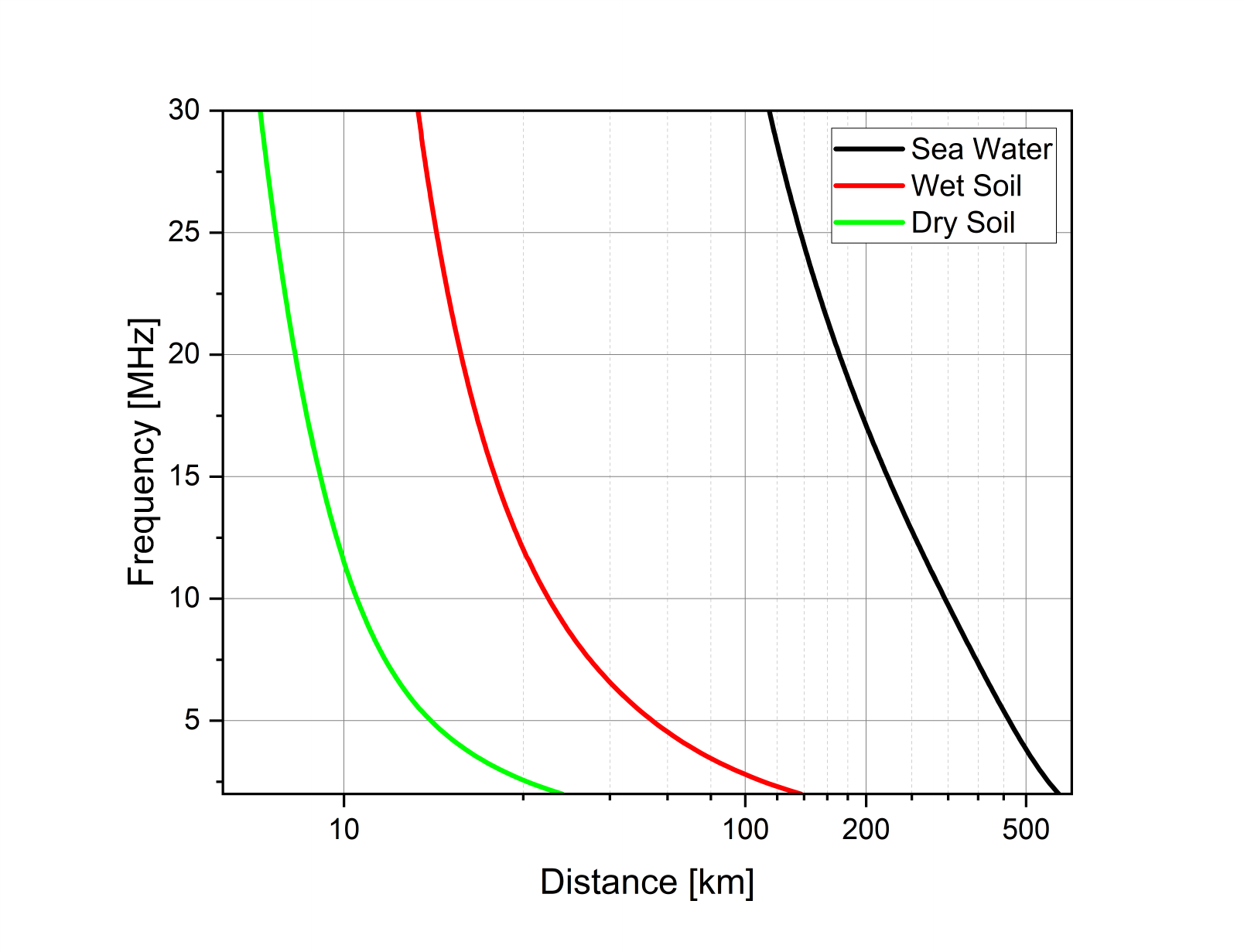}

    \caption{The relationship between frequency and propagation distance for ground waves for various ground types. Results were obtained using the LFMF Smooth Earth ground wave model \cite{NTIA_LFMF}. Ground conditions are as follows: Dry Soil ($\varepsilon_r=3, \sigma=0.001$ S/m), Wet Soil ($\varepsilon_r=25, \sigma=0.02$ S/m), Sea Water ($\varepsilon_r=80, \sigma=5$ S/m).}
    \label{fig:gw_f_vs_distance}
\end{figure}

While ground waves are capable of LOS links, it is often desirable to establish over the horizon (OTH) links. In this case, the surface wave is of greater importance. By placing the transmitter and receiver close to the ground (less than a wavelength), the reflection coefficient associated with the reflected wave tends to -1, leading to a cancellation of the direct and reflected waves. This also leads to a doubling of electric field intensity of the surface wave \cite{fabrizio_high_2013}. Bringing the antennas close to an imperfectly conducting surface, however, will affect the radiation pattern, which must be accounted for. \red{In practice, ground waves are suitable for distances between 200 km and 500 km, but this will depend on the transmit power, frequency, ground conductivity, permittivity, and the antenna heights off the ground \cite{dolukhanov_propagation_1995, fabrizio_high_2013}.} Fig. \ref{fig:gw_f_vs_distance} shows the ground wave propagation distance increase with decreasing frequency. Fig. \ref{fig:gw_f_vs_distance} also shows that surfaces with high conductivity, such as sea water, support longer communication distances. This is due to radiofrequency (RF) energy having a limited penetration depth in high conductivity materials \cite{balanis_balanis_2024, jin_theory_2010}. Therefore, as the conductivity of the ground decreases, more energy is lost in the ground and the wave is more rapidly attenuated. As far as polarization is concerned, horizontally polarized antennas are not practical due to severe attenuation of the electric field. Vertically polarized antennas are required for optimal performance.

\subsection{Unique Links}
Beyond the three traditional link types of  long range skywave, NVIS and ground wave, several circumstantial links exist. These consist of aurora, trans-equatorial, Pedersen Ray, F-layer long path, and gray line links.

Auroras stem from the solar flares and coronal mass ejections discussed in Section \ref{sec:links}-\ref{sec:weather}. Once the sun ejects these electrically charged particles, they travel to Earth interacting with Earth's magnetic field. Many of the particles are reflected off of the magnetic field and back into space, but some are funneled downward and collide with atoms and molecules in the atmosphere. This interaction excites the particles creating a wall of ionized particles. \red{These walls are called auroras and are capable of scattering high frequency transmissions \cite{silver_arrl_2022}.} By scattering the high frequency transmissions, long paths may be connected by positioning directive antennas towards the center of the aurora. According to \cite{silver_arrl_2022} paths as long as 2,300 km have been completed. Auroras have an inverse effect on low frequency transmission as they cause D-layer absorption to increase, thereby limiting the ability of low frequency transmissions to be completed.

\begin{figure}[h]
    \centering
    \includegraphics[width=1\columnwidth]{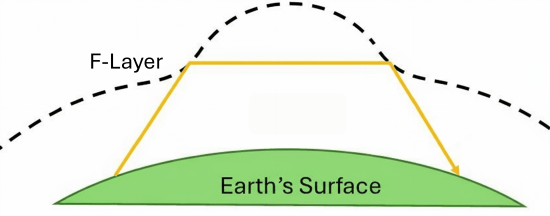}
    \caption{Trans-equatorial path, refracting off the slope of the ionosphere near the equator, and traversing to the opposite side of the equator to refract back to Earth.}
    \label{fig:TransEqui}
\end{figure}

Trans-equatorial paths are a phenomenon that occurs around the magnetic equator of Earth and support links varying from 5,000--8,000 km \cite{silver_arrl_2022}. To complete this link, both the transmitter and receiver must be placed equidistantly from the magnetic equator as to experience the same ionospheric characteristics. For trans-equatorial paths, the transmitted wave refracts off a slope in the ionosphere. This slope is a result of increased radiation exposure, forcing the F-layer to have a higher altitude and ion density locally around the magnetic equator \cite{silver_arrl_2022}. Once refracted off one side of the slope, the wave will propagate nearly parallel to Earth's surface until once again refracting off the same slope on the opposite side of the magnetic equator, directing the wave towards Earth's surface. Following this path, it can be noticed that the wave refracts off the ionosphere twice before returning to the Earth's surface. Fig. \ref{fig:TransEqui} provides a graphical representation of this path. \red{This mode is strongest near the spring and autumn equinoxes, favors late-afternoon and evening hours, and requires geomagnetically quiet conditions} \cite{silver_arrl_2022}.

The ARRL refers to the F-layer long path as the longer path around the Earth when making an F-layer communication \cite{silver_arrl_2022}. Because the connection depends on the circumference of the Earth, long paths in the F-layer are always more than 20,000 km, requiring multiple hops. Typically, these links will have lower SNR than the shorter F-layer path due to the number of hops and absorption resulting in losses of signal intensity. Nevertheless, in certain scenarios, the longer path can be beneficial. If the long path is over water and in the sun light, links at selective frequencies may be stronger or exist as compared to the shorter path. Additionally, the long path may also provide a feasible link if the short path is experiencing higher absorption or E-region blanketing due to ionospheric behaviors.

A time-specific form of propagation is gray line links. The gray line is the transition between nighttime and daytime. This region creates unusual ionospheric behaviors which can be utilized for long distance links \cite{silver_arrl_2022}. \red{Given that the F-layer is higher in the ionosphere, it will be within the sunlight, experiencing more solar radiation as opposed to the D- and E-layers which will still be within the shade of night, limiting their strength and therefore the absorption experienced within these layers.} By transmitting into this region, low frequency links, below 4 MHz can be created more than halfway around the Earth \cite{silver_arrl_2022}. The propagating wave is pushed from the gray line into the nighttime region of the sky, slowly returning to the gray line elsewhere on the Earth. This movement from the gray line to the nighttime and back is due to the tilt of the Earth. The wave is not refracted from the ionosphere back to Earth's surface until reaching the gray line again as it is likely ducted in the nighttime ionosphere. Ducting occurs when a wave is trapped in the ionosphere due to ion density irregularities allowing for long distance travel without refracting off of the Earth surface. More information on gray line propagation can be found in \cite{callaway_gray_nodate}.

\begin{figure}[h]
    \centering
    \includegraphics[width=1\columnwidth]{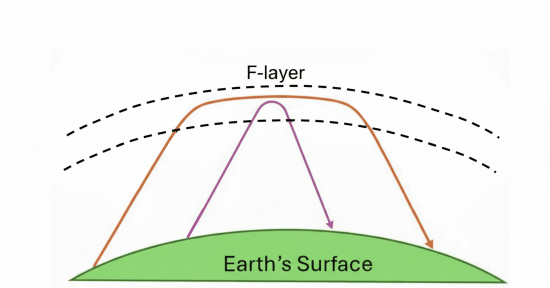}
    \caption{A Pedersen Ray, in orange, traveling much further in the ionosphere than a conventionally refracted wave, shown in purple. Propagation channeling through the ionosphere, as the Pedersen Ray allows for very long range communication, up to 12,000 km \cite{silver_arrl_2022}.}
    \label{fig:ped}
\end{figure}

Similar to ducting, Pedersen Rays or high-angle one-hop paths, are links where waves are incident at oblique angles very near those too great for transmission where the wave would escape into space \cite{silver_arrl_2022}. This incidence causes the ray to travel within the F2-layer of the ionosphere, following the contour of the Earth, as can be seen in Fig. \ref{fig:ped}. \red{By following this path within the ionosphere, Pedersen Rays allow for single-hop paths up to 12,000 km \cite{silver_arrl_2022}.} These links can be unreliable as they require stable ionospheric conditions but do offer the unique capability to create a single-hop low-loss long distance link.


\section{HF SKYWAVE PROPAGATION EXAMPLES}
\label{sec:examples}
In this section, we present several links which describe the effects of the ionosphere on radio wave propagation. Links are modeled using ionospheric modeling and ray tracing software, PHaRLAP \cite{DSTG_PHaRLAP, cervera2014}. Time of day and geographical location are considered. For comparison, two NVIS links are simulated, along with $\sim$2,000 km, $\sim$3,000 km, and $\sim$5,000 km long-range skywave links. An example of a ground wave link is provided in a previous section in Fig. \ref{fig:gw_f_vs_distance}. Ground wave link modeling is not dependent on local ionospheric conditions, nor for any specific transmitter/ receiver location, but solely on the ground conditions that exist between the transmitter and receiver, the heights of the antennas above the ground, and the antenna polarization.

\red{At the NVIS, 2,000~km, and 5,000~km distances, each comparison pairs a higher-latitude path with a near-equatorial path to emphasize the influence of ionospheric electron density, and therefore the usable frequency, on otherwise comparable path geometries. The near-equatorial paths possess a denser ionosphere, owing to greater solar exposure, than their higher-latitude counterparts. At 3,000~km, we instead pair two higher-latitude paths of nearly equal length but markedly different latitude, namely Orlando, Florida to Phoenix, Arizona (midpoint at $31^\circ$N) and Moscow, Russia to Barcelona, Spain (midpoint at $50^\circ$N), which isolates the effect of latitude under a fixed path geometry. Several of these links are subsequently corroborated against measured propagation data in Section~\ref{sec:examples}-\ref{subsec:Corroboration}.}


\begin{figure*}[!t]
    \centering
    \includegraphics[width=\textwidth]{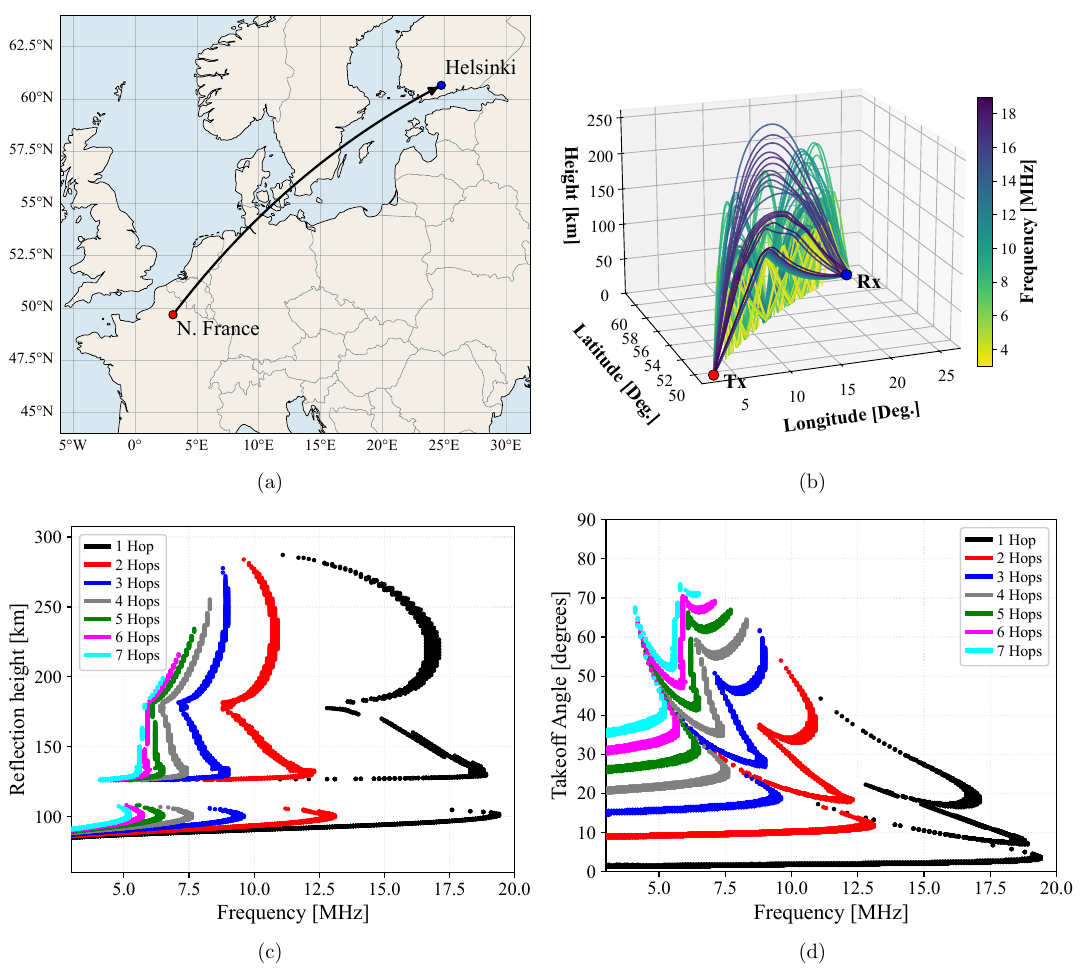}
    \caption{\red{Ray tracing data from PHaRLAP for the northern France to Helsinki, Finland link (daytime, 11:05~UT). (a) Great-circle map showing the transmitter and receiver and the 1{,}825~km path. (b) Simulated ray paths that reach the receiver, each colored by its frequency. (c) Oblique ionogram of reflection height versus frequency. (d) The same arrivals plotted as takeoff angle versus frequency.}}
    \label{fig:LayersAndHeights}
\end{figure*}

\subsection{Interpreting the Ionogram}
\label{InterpretingIonogram}

Ionograms, specifically oblique incidence ionograms, can be interpreted to predict link behavior. An ionogram is produced by an ionosonde (ionospheric sounder) and displays the refractive height as a function of sounding frequency. The ionogram reveals important information about a link, including the critical frequency and the height of a given ionospheric layer. A sample ionogram, simulated in PHaRLAP, is presented in Fig. \ref{fig:LayersAndHeights}. \red{This ionogram simulates the propagation of HF waves between northern France and Helsinki, Finland on July 15, 2024 at 11:05 UTC.} The ionospheric electron density profile is generated in PHaRLAP \red{using the International Reference Ionosphere (IRI-2020) model~\cite{bilitza2022}} by providing the time in UTC, the average sunspot number, and coordinates of the desired link. Fig. \ref{fig:LayersAndHeights}(a) shows the location of the transmitter and receiver sites on the map and Fig. \ref{fig:LayersAndHeights}(b) shows the 3D ray tracing results obtained from these simulations. By sweeping the azimuth, elevation, and frequency, \ref{fig:LayersAndHeights}(b) displays potential E- and F-Layer links. The results show the expected behavior of HF propagation where lower frequency links require higher TOAs and higher frequency links require lower TOAs to satisfy~(\ref{eqn:MUF}). \red{Links with up to seven hops were simulated.} Figs. \ref{fig:LayersAndHeights}(c) and \ref{fig:LayersAndHeights}(d), respectively, display the refractive heights as a function of frequency and the takeoff angle of the transmit antenna, needed to establish the desired link, as a function of frequency. The plotting style of Fig. \ref{fig:LayersAndHeights}(d) is adopted in this work to emphasize the practical consideration of antenna selection on link feasibility. The practical implementation of any HF system requires the TOA to match the radiation pattern of the antenna. Specifically,  to achieve the highest SNR for the link, the direction of maximum radiation of the transmit antenna should be towards the TOA.

By referencing Fig. \ref{fig:LayersAndHeights}(c) the HF refractive layers of the ionosphere are apparent. \red{The three refracting layers can be identified directly from the structure of the figure. The returns are organized into three distinct bands of reflection height, and within each band the reflection height rises gradually with frequency as the wave must penetrate closer to that layer's electron density peak before refracting. As the limiting frequency of a layer is approached, the trace turns sharply upward and the returns reappear in the next, higher band, where the same gradual rise repeats. These abrupt changes of direction, from a gentle upward slope to a near-vertical rise, occur twice with increasing altitude and mark the E- to F1- and F1- to F2-layer boundaries, delineating the separate band of frequencies and reflection heights that each layer occupies.} Now consider Fig. \ref{fig:LayersAndHeights}(d), where the same three bands and abrupt transitions appear in takeoff angle enabling the same layer identification techniques based on TOA. This analysis is applied throughout the remainder of this section and reveals important information about link establishment.

\red{For every link in this section, only rays arriving within a fixed receive radius of the destination are retained, and the maximum usable frequency is taken as the highest frequency supporting at least three such arrivals (to ensure that isolated single-ray returns do not set the band edge). A 50~km radius is used for the long-range skywave links; the NVIS links use a tighter 20~km radius, since at their short path lengths a 50~km radius would span an over-large fraction of the path and admit scattered, off-target returns.}

\begin{figure*}[!tb]
    \centering
    \includegraphics[width=\textwidth]{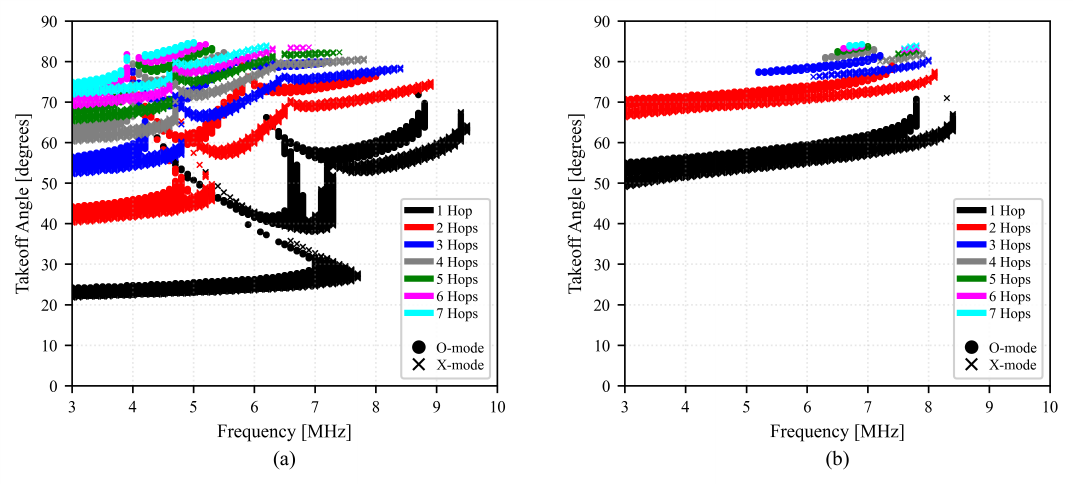}
    \caption{\red{(a) Daytime and (b) nighttime ionogram of a northern NVIS link from Frankfurt, Germany to Prague, Czech Republic. The number of refractions off the ionosphere is labeled as ``hops" and is delineated by the marker colors.}}
    \label{fig:FrankfurtPrague}
\end{figure*}

\begin{figure*}[!tb]
    \centering
    \includegraphics[width=\textwidth]{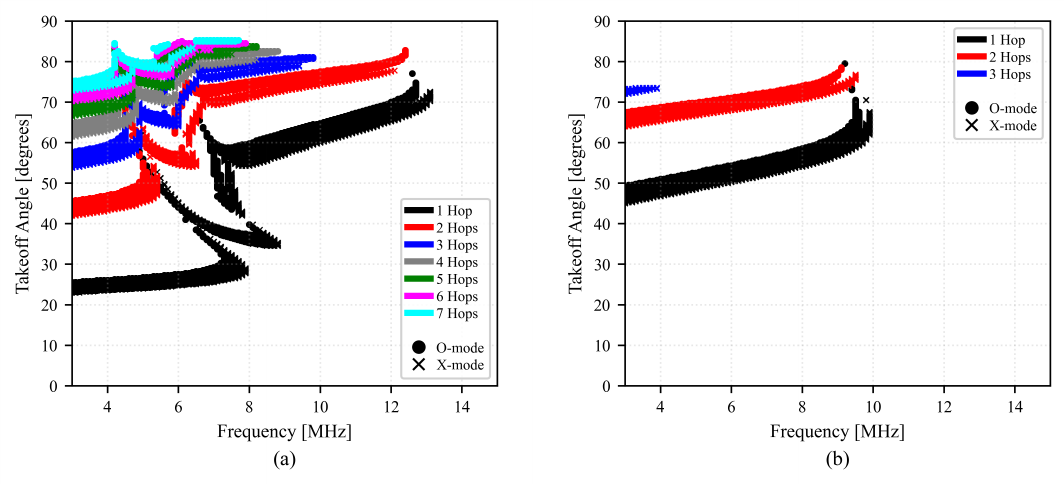}
    \caption{(a) Daytime and (b) nighttime ionogram of the southern NVIS link from San Jose, Guatemala, to Tegucigalpa, Honduras. Compared to the northern NVIS link, the maximum usable frequency (MUF) is higher during both day and night due to the increased electron density of the ionosphere near the equator.}
    \label{fig:SanJoseTeg}
\end{figure*}

\subsection{Representative NVIS Communication Links}

\red{The first comparison is of two NVIS paths: one spanning 411~km from Frankfurt, Germany to Prague, Czech Republic, and the second spanning 390~km from San Jose, Guatemala to Tegucigalpa, Honduras. July daytime and nighttime ionograms of these links are shown in Figs.~\ref{fig:FrankfurtPrague} and \ref{fig:SanJoseTeg}. The modeled daytime NVIS-MUF for the Frankfurt to Prague link is approximately 9.5~MHz, decreasing to roughly 8.4~MHz at night as the loss of solar illumination lowers the local electron density. As is characteristic of NVIS geometry, the link is established at high takeoff angles. In the daytime ionogram, Fig.~\ref{fig:FrankfurtPrague}(a), one-hop returns (black) span the band up to the NVIS-MUF, while multi-hop returns (two through seven hops) are confined to lower frequencies and progressively higher takeoff angles. The one-hop trace exhibits the characteristic NVIS structure, with lower-takeoff-angle E-layer returns and higher-angle F-layer returns; over an intermediate frequency interval the E- and F-layer one-hop solutions coexist, reflecting the continuous rather than sharply discrete layering discussed in Section~\ref{sec:examples}-\ref{InterpretingIonogram}.}

One-hop links are advantageous as they experience the least attenuation as they travel through the ionosphere. A one-hop F-layer ionospheric link encounters D- and E-layer absorption twice, once during ascent and once during descent, when the signal refracts from the F-layer. In contrast, multi-hop F-layer links experience greater attenuation due to repeated traversals through these lower ionospheric regions, as well as additional losses from ground reflections at the Earth's surface. Consequently, it is generally preferable to select a TOA that supports one-hop propagation to minimize overall path loss and maximize signal strength. \red{In Fig.~\ref{fig:FrankfurtPrague}(a), Frankfurt to Prague supports multi-hop links at the higher takeoff angles up to approximately 8~MHz.}

\red{At night, the modeled NVIS-MUF falls to approximately 8.4~MHz. This reduction results from the absence of solar radiation, which lowers the local electron density in the ionosphere. Similar to daytime conditions, multi-hop propagation remains possible at higher TOAs within the same frequency range as the one-hop link. However, E-layer refractions are largely absent at night because the E-layer dissipates after sunset.}

\red{Comparatively, the San Jose to Tegucigalpa link exhibits a higher MUF during both daytime and nighttime conditions, reaching approximately 13.1~MHz during the day and 9.9~MHz at night. This demonstrates the advantages of ionospheric paths located closer to the Earth's equator, where increased solar radiation enhances ionization and thereby raises the electron density of the ionosphere, supporting higher frequency refractions. The daytime usable frequency of the near-equatorial path is roughly 38\% higher than that of the higher-latitude Frankfurt to Prague path. This advantage is tempered by greater absorption within the denser lower ionospheric layers, which attenuate the signal during transmission. The magnitude of this attenuation is inversely proportional to frequency, making higher frequencies less susceptible to D- and E-layer absorption. Overall, comparison of the two links indicates that lower-latitude regions are generally more favorable for HF communications, as they enable a wider range of usable frequencies, specifically higher frequencies which experience less attenuation and noise.}

\begin{figure*}[!tb]
    \centering
    \includegraphics[width=\textwidth]{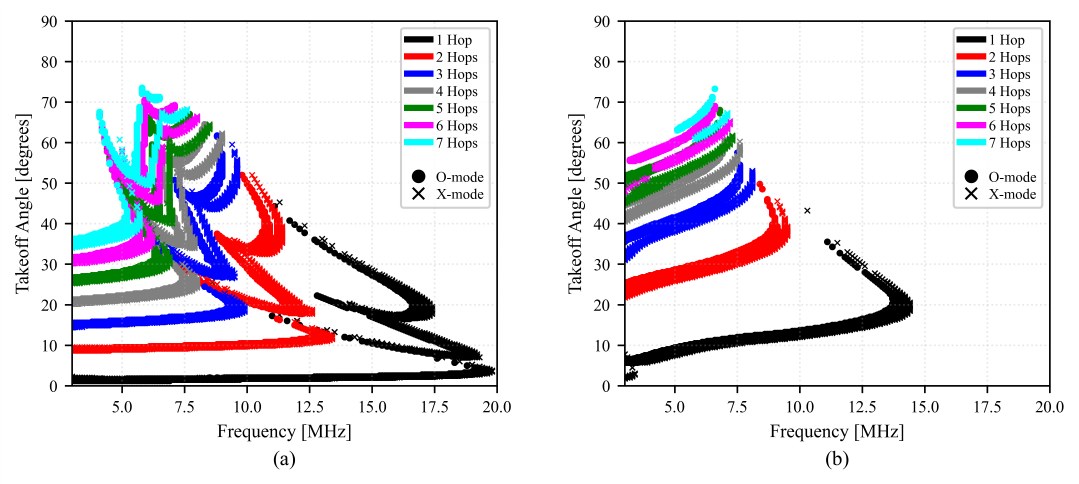}
    \caption{\red{(a) Daytime and (b) nighttime ionogram for the northern 2,000~km link from northern France to Helsinki, Finland.}}
    \label{fig:NFranceHelsinki}
\end{figure*}

\begin{figure*}[!tb]
    \centering
    \includegraphics[width=\textwidth]{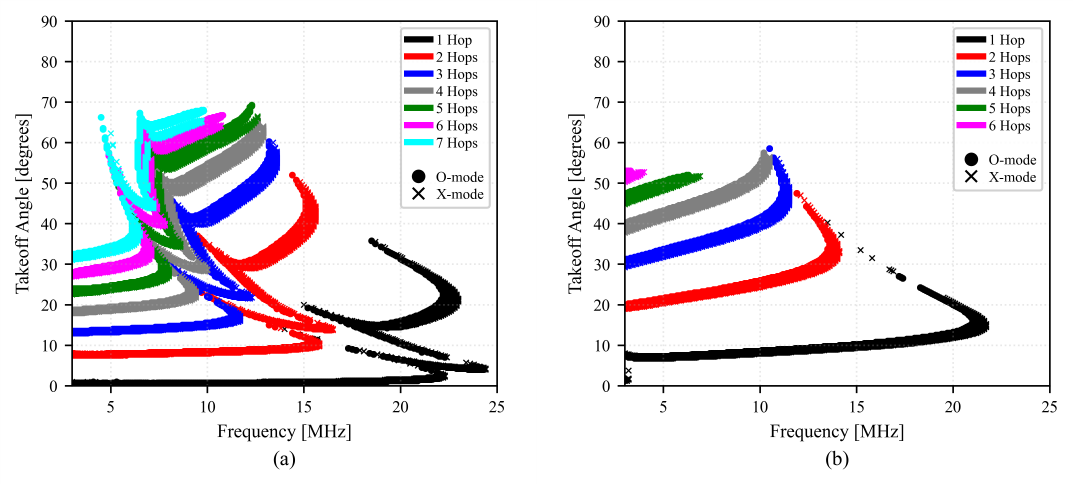}
    \caption{(a) Daytime and (b) nighttime ionogram for the southern 2,000 km link from San Jose, Guatemala to Port-au-Prince, Haiti.}
    \label{fig:SanJosePortAuPrince}
\end{figure*}

\subsection{Representative 2,000 km Long Range Skywave Links}

\red{A northern 2,000~km link was modeled between northern France and Helsinki, Finland; this is the same link used as the worked example in Fig.~\ref{fig:LayersAndHeights}. The corresponding oblique ionograms for both daytime and nighttime conditions in July of 2024 are shown in Fig.~\ref{fig:NFranceHelsinki}. This link utilizes a substantially higher portion of the HF spectrum than the NVIS links, with a modeled daytime MUF of approximately 19.8~MHz, decreasing to about 14.4~MHz at night. This increase relative to the NVIS case is a result of the lower incidence angle of the waves on the ionosphere, which allows higher frequencies to refract back to Earth. Examining the daytime ionogram, Fig.~\ref{fig:NFranceHelsinki}(a), one-hop returns (black) reach the MUF at low takeoff angles, on the order of $5^\circ$ at the upper frequencies, while two-hop returns (red) appear at higher takeoff angles; both E- and F-layer contributions are present. These low one-hop takeoff angles motivate the antenna considerations discussed next.}

\red{The practical difficulty of very low takeoff-angle propagation merits emphasis. Establishing a link with adequate SNR at such angles would require an antenna exhibiting high gain at very low elevation angles, which is generally difficult to achieve over terrestrial surfaces since most antennas exhibit a null at $0^\circ$ elevation when mounted on real earth. Furthermore, surrounding geographic features, such as terrain and buildings, would further restrict the viability of such low-angle radiation. Given an unconstrained resource environment, the desired low TOA could be achieved using the aforementioned LPDA by mounting it on an elevated tower. This configuration minimizes ground effects on radiation patterns, ensures clearance over surrounding topography, and facilitates mechanical tilting to adjust the elevation pattern. Other antenna designs can be similarly positionally optimized, though their specific setup would be determined by the antenna's radiation pattern and environment.}

\red{During nighttime conditions, Fig.~\ref{fig:NFranceHelsinki}(b), the takeoff angles associated with different hop counts are well separated. For example, near 5~MHz the one-, two-, and three-hop paths arrive at takeoff angles of approximately $10^\circ$, $20^\circ$, and $30^\circ$, respectively. Because these angles are well separated, a directive transmit antenna can be designed with maximum gain toward one hop count while exhibiting substantially lower gain toward the others, so that the corresponding path dominates at the receiver. Each path also introduces a distinct propagation delay, which can produce multipath interference; however, because the favored path arrives with substantially higher power, the impact of the weaker paths can be significantly lower.}

\red{Comparing this 2,000~km northern link to the corresponding southern 2,000~km link reveals several similarities, including daytime E- and F-layer one- and two-hop propagation and distinct nighttime TOAs corresponding to different hop counts. The primary difference lies in the MUF. As shown in Fig.~\ref{fig:SanJosePortAuPrince}(a), the modeled daytime MUF of the southern San Jose to Port-au-Prince link is approximately 24.4~MHz, roughly 4.6~MHz higher than that of the northern link. This enhancement results from the higher electron density present in the equatorial path. The nighttime ionogram for the San Jose to Port-au-Prince path likewise shows a substantially higher MUF (approximately 21.7~MHz) than the northern link. Overall, comparison of the two 2,000~km links highlights the advantages of equatorial propagation, where elevated ionization levels enable higher MUFs in both day and night conditions, thereby expanding the usable bandwidth of the HF spectrum.}

\begin{figure*}[!tb]
    \centering
    \includegraphics[width=\textwidth]{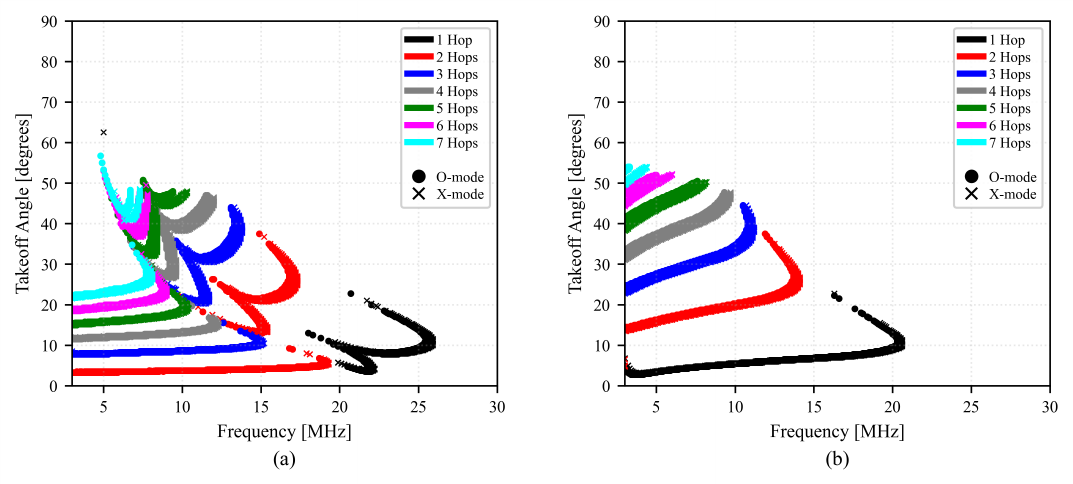}
    \caption{\red{(a) Daytime and (b) nighttime ionogram for the 3,000~km link from Orlando, Florida ($31^\circ$N) to Phoenix, Arizona.}}
    \label{fig:OrlandoPhoenix}
\end{figure*}
\begin{figure*}[!tb]
    \centering
    \includegraphics[width=\textwidth]{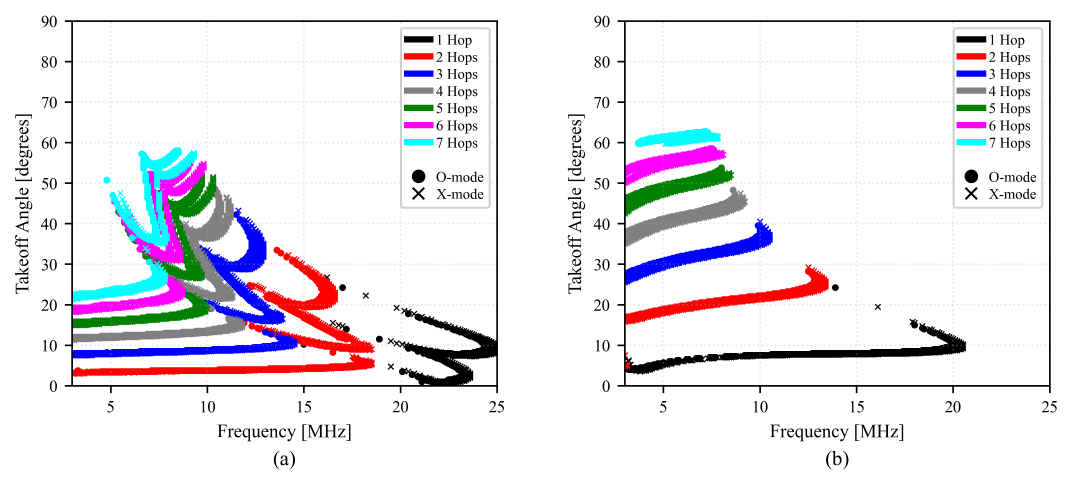}
    \caption{\red{(a) Daytime and (b) nighttime ionogram for the 3,000~km link from Moscow, Russia ($50^\circ$N) to Barcelona, Spain.}}
    \label{fig:MoscowBarcelona}
\end{figure*}

\subsection{Representative 3,000 km Long Range Skywave Links}
\red{At 3,000~km we depart from the higher-versus-lower-latitude pairing used at the other distances and instead compare two higher-latitude paths of nearly identical length but different latitude: Orlando, Florida to Phoenix, Arizona, and Moscow, Russia to Barcelona, Spain, with path-midpoint latitudes near $31^\circ$N and $50^\circ$N, respectively. Holding the path geometry essentially fixed in this way isolates the influence of latitude on the modeled MUF. The corresponding ionograms are shown in Figs.~\ref{fig:OrlandoPhoenix} and \ref{fig:MoscowBarcelona}.}

\red{The modeled MUFs of the two links are remarkably close despite the roughly $19^\circ$ difference in latitude. The Orlando to Phoenix link has a daytime MUF of approximately 25.9~MHz and a nighttime MUF of 20.6~MHz, while the Moscow to Barcelona link reaches approximately 25.0~MHz during the day and 20.5~MHz at night. In July, the higher-latitude path benefits from extended summer daylight at these latitudes, which offsets much of the electron-density advantage that the lower-latitude path would otherwise enjoy. Thus, the two links present a similar usable-frequency ceiling at this distance. Both links lie well within the long-range skywave regime, in which the lower incidence angle on the ionosphere raises the MUF relative to the shorter links examined above. Examining the daytime ionograms, Figs.~\ref{fig:OrlandoPhoenix}(a) and \ref{fig:MoscowBarcelona}(a), both links support one-hop F-layer propagation at low takeoff angles of $\sim10^\circ$ near the MUF together with a sequence of multi-hop returns at progressively higher takeoff angles. Single-hop E-layer propagation is not supported at this distance, as a 3,000~km E-layer hop would require an unrealistically low, near-grazing takeoff angle. The same hop structure persists at night, Figs.~\ref{fig:OrlandoPhoenix}(b) and \ref{fig:MoscowBarcelona}(b), at the reduced nighttime MUF. That the two links exhibit closely matching hop structures as well as nearly coincident MUFs, despite their latitude difference, reinforces the point above.}

\red{Because these two links are nearly co-located in distance but well separated in latitude, and because both fall within FT8-dense corridors with a nearby ionosonde, they form the centerpiece of the measured corroboration presented in Section~\ref{sec:examples}-\ref{subsec:Corroboration}, where the modeled MUFs above are compared against crowdsourced FT8 observations and direct ionosonde measurements.}

\begin{figure*}[!tb]
    \centering
    \includegraphics[width=\textwidth]{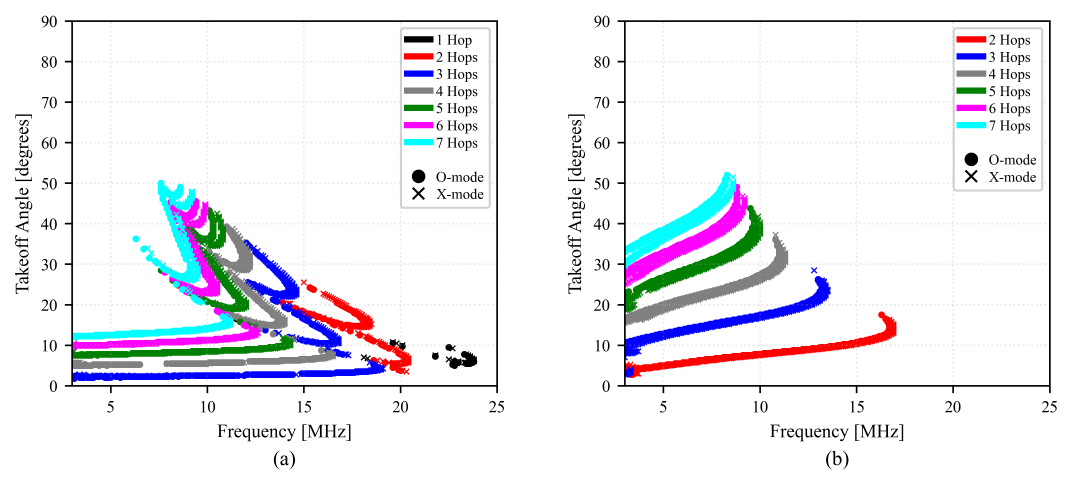}
    \caption{\red{(a) Daytime and (b) nighttime ionogram for the northern 5,000~km transatlantic link from Newark, New Jersey to Glasgow, Scotland.}}
    \label{fig:NewarkGlasgow}
\end{figure*}

\subsection{Representative 5,000 km Long Range Skywave Links}
\red{The final examples considered are the 5,000~km links. For the higher-latitude case, a transatlantic path from Newark, New Jersey, to Glasgow, Scotland, was selected. Continuing the observed trend of increasing MUF with greater path length, the modeled daytime MUF of this link is approximately 23.8~MHz, with a nighttime MUF of approximately 16.9~MHz. Examining the daytime ionogram, Fig.~\ref{fig:NewarkGlasgow}(a), propagation is predominantly multi-hop; at most a sparse one-hop F-layer trace appears at low takeoff angle, and single-hop E-layer propagation is excluded by the path length. At night, Fig.~\ref{fig:NewarkGlasgow}(b), one-hop propagation is no longer supported at all (the lowest realizable mode is two hops) since the reduced electron density cannot return a single 5,000~km hop. As a transatlantic, multi-hop F-layer path, this link is corroborated against crowdsourced FT8 observations in Section~\ref{sec:examples}-\ref{subsec:Corroboration}. Its transatlantic geometry has no single-station ionosonde reference. Thus, it is corroborated by the FT8 operability ceiling alone.}

\begin{figure*}[!tb]
    \centering
    \includegraphics[width=\textwidth]{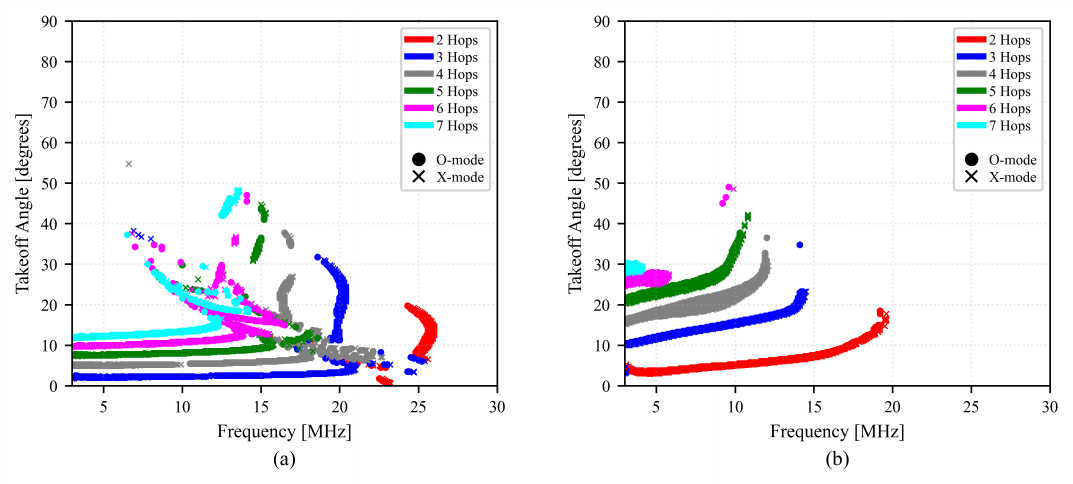}
    \caption{\red{(a) Daytime and (b) nighttime ionogram for the southern 5,000~km link from San Juan, Puerto Rico to Dakar, Senegal.}}
    \label{fig:SanJuanDakar}
\end{figure*}

Spanning the entire Atlantic Ocean, the southern link from San Juan, Puerto Rico to Dakar, Senegal, exhibits the same characteristic increase in MUF. This link enables coverage across nearly the entire HF band.  \red{Compared with the higher-latitude 5,000~km Newark to Glasgow link, the San Juan to Dakar link presents a higher MUF in both day and night ($\approx$26.0~MHz versus 23.8~MHz during the day and 19.6~MHz versus 16.9~MHz at night) consistent with the denser equatorial ionosphere.} This again demonstrates the practical upper limit of HF communications. As the paths begin to reach longer distances, the minimum number of hops supporting a viable link begins to rise.

Another clear feature in the daytime ionogram is the distinct separation between E-layer and F-layer refraction paths for each hop count. As described previously, such differentiation allows for an informed inference regarding the refracting layer responsible for each signal path. In this case, the F-layer refractions are clearly delineated from the E-layer paths, demonstrating that even for seven-hop propagation, F-layer refraction is not supported below approximately 6 MHz. The E-layer refracted links also occur at discrete intervals of TOA corresponding to hop number. However, largely consistent with the higher-latitude 5,000 km case, the E-layer does not support links with fewer than three hops (except for a small bandwidth of two-hops from 22 - 24 MHz) due to the excessive path length, which would require unrealistically low (negative) TOAs.

Through the analysis and comparison of NVIS, 2,000 km, 3,000 km, and 5,000 km long-range skywave links, it becomes evident that lower latitude (southern) locations support higher usable frequencies during both daytime and nighttime conditions. This advantage arises from the increased electron density present in the equatorial ionosphere. However, southern links also experience greater absorption within the D- and E-layers. The comparison further provides valuable insight into the TOAs required for various path lengths; as the link distance increases, the necessary TOA decreases, eventually reaching values that are physically unrealistic, thereby constraining the feasibility of long single-hop propagation paths.

\subsection{Corroboration of the Simulated MUF Against Measured Propagation Data}
\label{subsec:Corroboration}

\red{The simulated maximum usable frequencies presented above are corroborated here against two independent sets of real propagation observations spanning the same July 2024 epoch. These are the operating-frequency ceiling inferred from crowdsourced amateur-radio (FT8) reception reports, and the MUF implied by direct ionosonde measurements of the F2-layer critical frequency. Five of the eight links fall within regions of sufficient observational coverage to support this comparison; the three near-equatorial links lack such coverage and remain model-only.}

\red{FT8 is a widely used amateur-radio digital mode based on 8-tone frequency-shift keying, in which stations exchange short, structured messages in 15-second transmit and receive cycles and successful decodes are automatically logged, with time stamps, frequencies, and station locations, to public reporting networks~\cite{franke2020}. The FT8 observable is a calibration-free operability ceiling. For each link, FT8 reception reports distributed through the CEDAR Madrigal database~\cite{madrigal_db} are collected between transmit and receive regions centered on data-driven hub centroids. These hubs are dense clusters of amateur activity discovered from the spot data itself, which also serve as the modeled path endpoints. For each hour the highest amateur band carrying a meaningful share (at least 2\%) of that hour's reports is identified. The upper edge of this occupied passband is the operability ceiling. Because it reflects only bands on which operators were active and successfully decoded, the ceiling is a strict lower bound on the true MUF; the simulated and ionosonde MUFs meeting or exceeding it is therefore the physically consistent outcome. The method used to process the CEDAR Madrigal measurement data requires no power calibration and is detailed in the supplementary material accompanying this paper.}

\red{The ionosonde observable is obtained from GIRO (Global Ionospheric Radio Observatory) ionosonde measurements~\cite{reinisch2011giro} of the F2-layer ordinary-wave critical frequency, $f_{o\textrm{F2}}$, on or near each path. The oblique-path MUF is recovered by scaling the measured vertical critical frequency by an obliquity factor (OF), $\mathrm{MUF}=f_{o\textrm{F2}}\times\mathrm{OF}$, which accounts for the oblique incidence geometry of the link. The obliquity factor is computed for each link from the mirror geometry of a single F2-layer hop, using half the great-circle range and the F2-layer reflection height to obtain the incidence angle on the layer; it is the secant of that incidence angle, equivalent to the $1/\sin\theta$ factor in (\ref{eqn:MUF}). At the NVIS and 2,000~km distances, the FT8 operability ceiling in July is set by sporadic-E propagation~\cite{whitehead1989} (roughly 21--29~MHz) rather than by the F-layer MUF. Therefore, these two links are corroborated by the ionosonde measurements, which probe the critical frequency of the F-layer directly.}

\begin{figure*}[t]
    \centering
    \includegraphics[width=\textwidth]{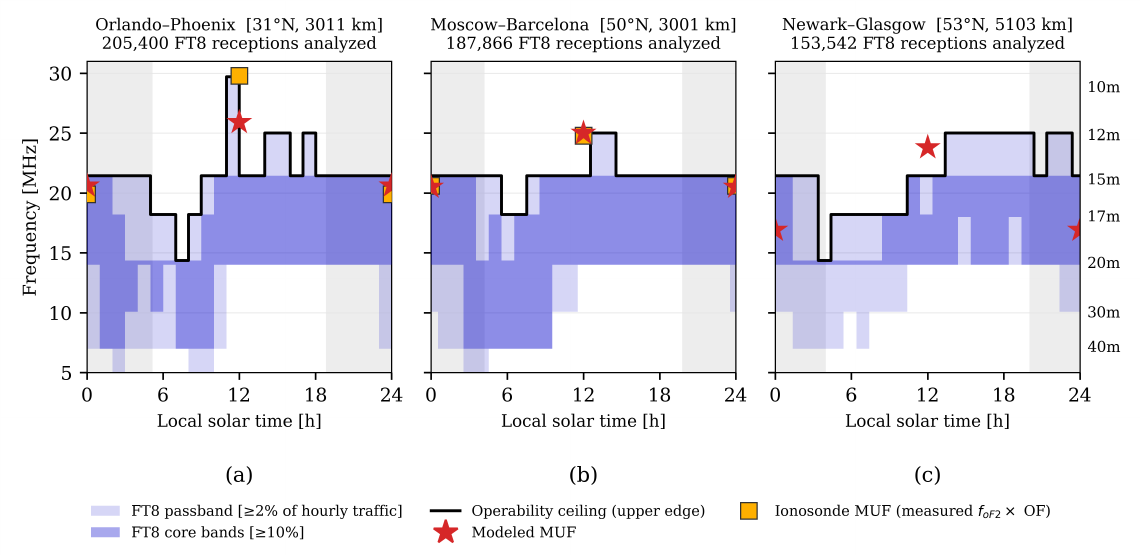}
    \caption{\red{FT8 operability ceiling compared with the simulated and ionosonde MUFs for the three longest corroborated links. Blue fill: FT8 passband; solid black line: operability ceiling; red stars: simulated MUF; amber squares: ionosonde MUF; gray shaded spans: night at path midpoint. Each panel's title gives the link endpoints, path-midpoint latitude, distance, and the number of FT8 receptions analyzed.}}
    \label{fig:corroboration}
\end{figure*}

\red{Fig.~\ref{fig:corroboration} presents this comparison for the three longest corroborated links, where multi-hop F-layer propagation dominates the path and the FT8 ceiling is a genuine F-layer observable. The FT8 passband and ceiling shown there are derived from the crowdsourced FT8 spot reports in the CEDAR Madrigal database~\cite{madrigal_db}, aggregated by hour over July~12--18, 2024 and over both propagation directions for each link (e.g., Frankfurt to Prague and Prague to Frankfurt). The week-long window is used because the FT8 operability ceiling requires sufficient decode density per hour--band--corridor bin to be statistically meaningful, whereas the IRI-based model, being a monthly-median climatology with no day-to-day variability, is evaluated at a single representative mid-window date (15 July 2024). The light-blue fill marks bands that carry at least 2\% of that hour's receptions, the darker fill marks core bands that carry at least 10\% of the transmissions, and the black operability ceiling traces the upper edge of the 2\% passband. The red stars are the simulated MUF at local noon and midnight, the amber squares are the measured ionosonde MUF multiplied by the obliquity factor, and the shaded spans denote night as delimited by the geometric solar terminator at the path-midpoint latitude for 15~July~2024. Across all three links, the simulated MUF tracks the measured ceiling through the diurnal cycle, sitting at or above it as expected for a lower bound.}

\begin{table*}[t]
\centering
\caption{Simulated MUF compared with the measured FT8 operability ceiling and the ionosonde-derived measured MUF ($f_{o\textrm{F2}}\times\,OF$) for the five corroborated links. All values are in MHz; ``---'' denotes an instrument that is not applicable to that link.}
\label{tab:corroboration}
{\color{black}
\begin{tabular}{l c c c c c c c}
\toprule
 & & \multicolumn{2}{c}{Simulated} & \multicolumn{2}{c}{Meas.\ FT8 Ceiling} & \multicolumn{2}{c}{Measured $f_{o\textrm{F2}}\times$\,OF} \\
\cmidrule(lr){3-4}\cmidrule(lr){5-6}\cmidrule(lr){7-8}
Link & Dist.\ (km) & Day & Night & Day & Night & Day & Night \\
\midrule
Frankfurt--Prague (NVIS) & 411  & 9.5  & 8.4  & ---  & ---  & 8.9  & 8.2  \\
N.\ France--Helsinki     & 1825 & 19.8 & 14.4 & ---  & ---  & 19.6 & 16.2 \\
Orlando--Phoenix         & 3011 & 25.9 & 20.6 & 28.9 & 21.2 & 29.8 & 19.9 \\
Moscow--Barcelona        & 3001 & 25.0 & 20.5 & 24.9 & 21.2 & 24.8 & 20.6 \\
Newark--Glasgow          & 5103 & 23.8 & 16.9 & 24.9 & 21.2 & ---  & ---  \\
\bottomrule
\end{tabular}
}

\vspace{2pt}
{\footnotesize\red{OF: obliquity factor, which scales the measured vertical critical frequency $f_{o\textrm{F2}}$ to the oblique-path MUF.}}
\end{table*}

\red{Table~\ref{tab:corroboration} summarizes the comparison. The simulated MUFs are the robust-edge values defined in Section~\ref{sec:examples} (a 50~km receive radius, or 20~km for the NVIS links). In the FT8 columns, the Day entry is the daytime-maximum ceiling and the Night entry is the local-midnight ceiling. The FT8 columns are left blank for the NVIS and 2{,}000~km links, where the July operability ceiling is set by sporadic-E rather than the F-layer MUF, and the ionosonde column is blank for Newark--Glasgow, a two-hop path with no single-station $f_{o\textrm{F2}}$ reference. Agreement is closest for the compact, uniform-latitude paths: for Moscow--Barcelona the simulated, FT8, and ionosonde values agree to within a few tenths of a megahertz in both day and night. Three deviations have clear physical explanations. Orlando--Phoenix crosses the low-latitude region of enhanced ionization, where both the FT8 noon peak ($\approx$28.9~MHz) and the ionosonde ($\approx$29.8~MHz) exceed the ray-traced daytime MUF (25.9~MHz); because the two independent observations agree with each other, the simulation is the conservative estimate in this case. For Newark--Glasgow, the nighttime FT8 ceiling exceeds the simulated value because the MUF of this 5{,}000~km path is set by its most-sunlit control point, whereas the simulation is evaluated at the path midpoint at local midnight, making the midpoint-midnight value a lower estimate. For N.~France--Helsinki, the nighttime ionosonde MUF exceeds the simulation by roughly 1.8~MHz, consistent with the path extending into the high-latitude ionospheric trough near $61^\circ$N, where the simulated electron density is comparatively low.}

\red{Taken together, the comparison corroborates the simulated MUFs across the $31$--$56^\circ$N latitude span of the five links using two physically independent observables. The remaining discrepancies are not random scatter but follow known ionospheric structure, namely the low-latitude enhancement and the high-latitude trough, as well as the geometry of multi-hop paths; in every case the simulated MUF is the conservative estimate relative to the measurements.}

\subsection{Link Budget Analysis}

\begin{figure*}[!t]
    \centering
    \includegraphics[width=\textwidth]{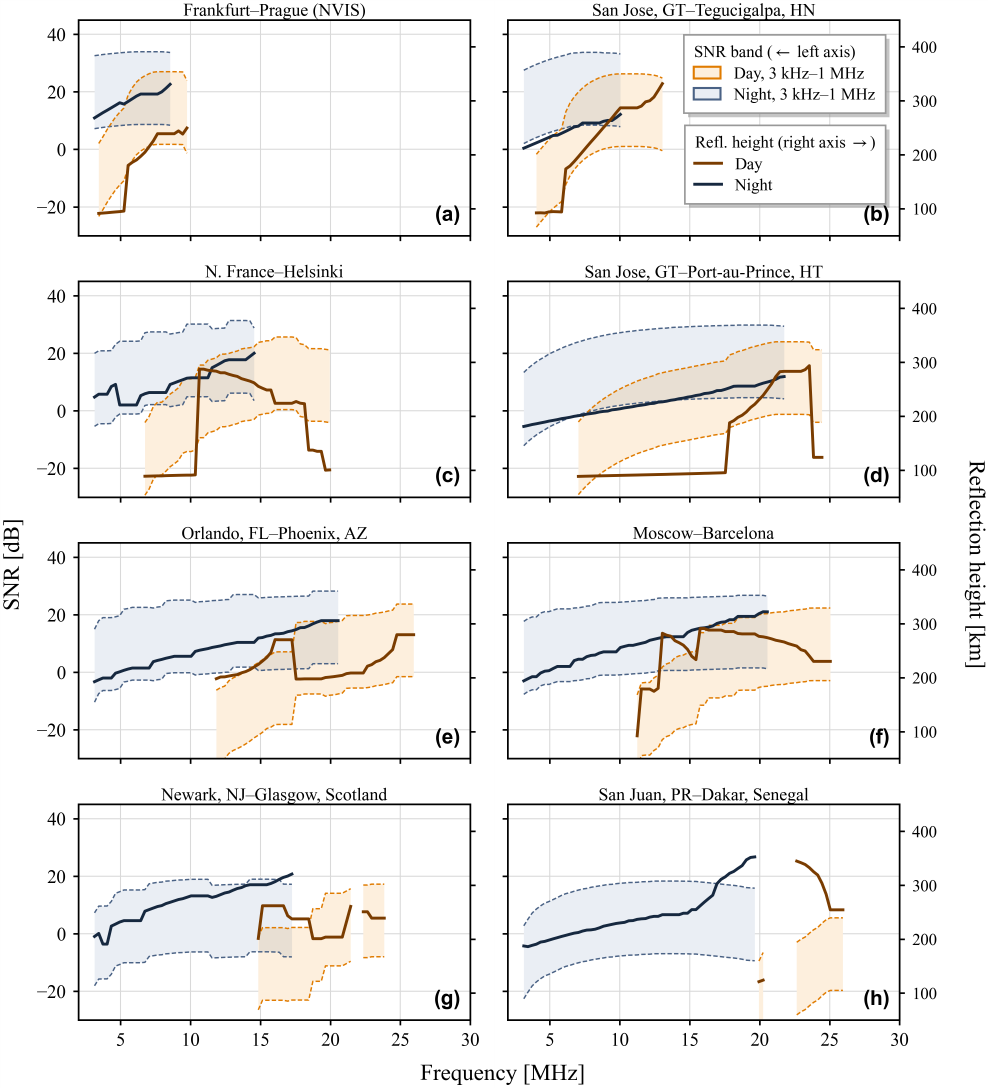}
    \caption{\red{Wideband HF skywave SNR for the eight study links (NVIS to 5{,}000~km), daytime and nighttime. In each panel, the shaded region is the received SNR across signal bandwidths from 3~kHz (upper edge) to 1~MHz (lower edge); the four intermediate bandwidths lie within the band, offset by $-10\log_{10}(\mathrm{BW}/3~\mathrm{kHz})$. Solid curves give the reflection (apogee) height of the strongest supported arrival at each frequency. SNR bands are read from the left axis and reflection-height curves from the right axis, as indicated by the legend. Model: 1~kW isotropic, IRI-2020 (R12~=~155, 15~July~2024), with free-space, absorption, ground, and polarization losses and combined atmospheric, galactic, and worst-case urban man-made noise per ITU-R~P.372.}}
    \label{fig:linkbudget}
\end{figure*}

\red{Fig.~\ref{fig:linkbudget} presents a wideband link-budget analysis for the eight study links, and Table~\ref{tab:lb-nfrance-helsinki} in the Appendix gives the corresponding tabulated budget for one representative link (northern France--Helsinki) so that the per-frequency values underlying the figure can be audited. \red{The link-budget analysis was performed using a custom simulation and post-processing pipeline that pairs PHaRLAP ionospheric ray tracing~\cite{DSTG_PHaRLAP} with the noise model of ITU-R Recommendation P.372~\cite{noauthor_recommendation_nodate-1}.} For each link, a 3D ray-tracing simulation was performed across the entire HF band (3--30 MHz) with a 1 kW transmit power. From the link budget equation~(\ref{eqn:lba}), the antenna gains $G_{tx}$ and $G_{rx}$ are assumed to be 0 dB (isotropic radiator). \red{Per-arrival free-space, ionospheric absorption, ground-reflection, and polarization-mismatch losses were included, and the atmospheric, galactic, and man-made noise contributions were combined following the full statistical procedure of ITU-R~P.372 \cite{noauthor_recommendation_nodate-1}. Man-made noise was set to the worst-case urban (``business'') category uniformly across all links, a conservative assumption adopted because the receiver environments are mixed.} Insertion losses due to feed lines, connectors, or additional RF components such as filters or couplers are not considered.}

\red{In both the figure and the table, results are separated by time of day, and each plotted point or table row is the strongest supported arrival at that frequency, reported with the reflection (apogee) height that PHaRLAP returns natively.} The received signal power, $P_{rx}$, is reported in dBm. The receive noise power is dependent on the receiver bandwidth and is therefore not reported in the table. Noise power, $P_n$, in dBm, for a given bandwidth can be calculated using $P_n = P_{rx} - \mathrm{SNR}$. \red{The table lists SNR at 3~kHz, 50~kHz, 100~kHz, 250~kHz, 500~kHz, and 1~MHz; in the figure these six bandwidths correspond to the shaded band, whose upper edge is the 3~kHz SNR and whose lower edge is the 1~MHz SNR.} \red{The SNR bands and reflection-height curves in Fig.~\ref{fig:linkbudget} are drawn only over frequencies at which the link is supported, defined as the strongest arrival achieving an SNR of at least $-10$~dB in the 3~kHz reference bandwidth; the plotted band is further clipped to robust lower and upper edges obtained with the same at-least-three-arrival criterion used for the reported MUFs in Section~\ref{sec:examples}.}

\red{The figure reveals the physics of ionospheric links in action, showing the effects of link distance, type (NVIS versus long-range skywave), time of day, and bandwidth on the received SNR. Consider the Frankfurt--Prague NVIS link (Fig.~\ref{fig:linkbudget}(a)), a mid-latitude short-range link characterized by high takeoff angles and low usable frequencies. During the day its low-frequency band suffers poor SNR owing to the high atmospheric noise and D-region absorption at those frequencies; at night the band shifts lower in frequency and improves as the absorbing D-region disappears. The equatorial San Juan--Dakar link (Fig.~\ref{fig:linkbudget}(h)) shows the opposite extreme. Although the daytime ray tracing supports propagation across nearly the entire HF band (Fig.~\ref{fig:SanJuanDakar}(a)), the daytime link budget confines the usable band to a relatively narrow window near the top of the HF spectrum. Over this 5{,}000~km multi-hop path the waves transit the sunlit D-region on every hop, and the cumulative absorption, which grows rapidly toward lower frequencies, suppresses the SNR of the lower bands below usable levels. Within this daytime window the strongest return reflects near the F2 peak before handing over to a lower reflection height near the junction frequency (approximately 23~MHz). At night the absorbing D-region disappears and the usable band widens and shifts downward in frequency, while the reduced electron density lowers its upper edge. Table~\ref{tab:lb-nfrance-helsinki} lists the corresponding per-frequency budget for the northern France--Helsinki link, giving the received power, reflection height, takeoff angle, and SNR at each of the six bandwidths for both day and night.}

Examining SNR as a function of bandwidth, we see a similar trend among all links. SNR drops rapidly as a function of increasing bandwidth. This is primarily due to the increase in noise power at wider bandwidths as is seen in~(\ref{eqn:noise_power}). Going from a 3 kHz bandwidth to a 1 MHz bandwidth, there is an approximate 25 dB increase in noise power, or equivalently, there is a 25 dB decrease in SNR. Of course, link quality can be improved by selecting adequate antennas with high gain and, optionally, wide instantaneous bandwidths.

\section{DISCUSSION}
HF communication represents a unique form of long-range transmission that utilizes the ionosphere as a refractive medium, enabling signal propagation over short and long distances through NVIS and long-range skywave propagation. \red{Though once used as a primary method of long-distance wireless communications, HF is now primarily used for defense and emergency communications when other forms, such as satellite communications, are unavailable.} Additionally, there exist many amateur radio enthusiasts who utilize the band for Ham radio communications. Due to the unpredictable nature of the ionosphere, a given HF link may be characterized by rapid changes in link geometry, fading and interference, and variability in frequencies supporting reliable communications. The affinity for HF communications is the ease of deployment of a transmit-receive site and relatively low-cost of operation. The large operating wavelengths allow for waves to be refracted off the ionosphere and travel large distances, supporting extremely long range BLOS communications with minimal infrastructure requirements.

HF communications support three distinct modes: ground waves, NVIS, and long range skywave. Ground wave supports short distance communications up to 500 km including line-of-sight and over-the-horizon links. Long-range skywave utilizes the ionosphere as a refractive medium to support communication distances in excess of 40,000 km. NVIS links are used to cover the \emph{skip zone} that exists between ground wave and long range skywave and are accomplished using high takeoff angles. The examples presented in Section \ref{sec:examples} demonstrate that lower TOAs support longer propagation distances by lowering the effective angle of incidence, thereby increasing the MUF, as described by (\ref{eqn:MUF}). \red{For the five links with sufficient observational coverage, the simulated MUFs were further corroborated against two independent sets of measured propagation data, the crowdsourced FT8 operability ceiling and direct ionosonde measurements, with the simulation proving to be the conservative estimate in every case and the remaining deviations following known ionospheric structure.} \red{It is also clear that transmitting at lower latitudes, near the Earth's geomagnetic equator, supports transmissions at higher frequencies because of the increased electron densities there, and these higher frequencies are in turn less susceptible to noise. The caveat is increased attenuation from the denser D- and E-layers, a loss that can be offset in the link budget by using highly directive antennas.}

A literature review of wideband HF experiments was presented in Section \ref{sec:bandwidth}-\ref{sec:wideband}. While successful wideband transmissions have been reported, their practical use remains constrained by the narrow bandwidth allocations within the HF spectrum, the variability of ionospheric conditions and the corresponding limitations on achievable data rates. The review, simulations, and hardware considerations presented here aim to guide in the development of wideband HF systems. Although modern systems have largely supplanted HF, reliable fallback communication methods remain essential during emergencies and infrastructure failures. The ease of deployment of transmit-receive HF systems make them opportune for such scenarios.

To facilitate wideband HF transmissions, appropriate hardware selection, namely antenna type, orientation, and site selection, must be properly vetted. In Section \ref{sec:antennas}, a review of popular radiation systems for HF communications was presented. LPDAs, folded dipoles, and rhombic antennas are good examples of HF antennas that support multi-octave frequency operation. In wideband HF applications, LPDAs appear to be particularly attractive for long range skywave and TTFDs are a good choice for NVIS applications. For wideband ground wave communications, antenna selection invokes tradeoffs between size and radiation efficiency. Wideband vertically polarized antennas placed near a good conducting ground are the element of choice in this operational environment. Even then, usable frequencies for ground waves are limited unless the ground wave link is used for communication over seawater (e.g., ship to shore or island to mainland links). Experimental investigations of wideband HF links should likely focus on skywave links initially. As for geographical location, an open, obstruction-free area, in a rural setting near the equator will maximize the success of wideband communications. \red{Finally, emerging antenna technologies may relax the size and bandwidth constraints that conventional designs impose on wideband HF systems. In particular, non-LTI electrically small HF antennas have recently demonstrated over-the-air transmission of wideband, digitally modulated HF waveforms from compact radiators~\cite{ma2025overair, ludois2026qmod}, and may enable far more compact wideband HF transmitters than the large broadband structures discussed above.} 

\appendix
\label{app:link_budgets}

\red{The wideband link-budget analysis for all eight study links is presented as a
summary figure in Section~\ref{sec:examples} (Fig.~\ref{fig:linkbudget}). A single
representative link-budget table is retained here, for the northern France--Helsinki
link, so that the per-frequency received power, reflection height, takeoff angle,
and bandwidth-dependent SNR underlying the figure can be directly verified.}

\begin{table*}[t]
\centering
\caption{\red{Northern France - Helsinki, Finland link budget analysis.}}
\label{tab:lb-nfrance-helsinki}
\begin{tabular}{ll c c c c c c c c c}
\hline
 & Freq. (MHz) & Refl. ht (km) & TOA ($^\circ$) & $P_{rx}$ (dBm) & \multicolumn{6}{c}{SNR (dB) vs Bandwidth} \\
\cline{6-11}
 & & & & & 3 kHz & 50 kHz & 100 kHz & 250 kHz & 500 kHz & 1000 kHz \\
\hline
\multirow{5}{*}{\textit{Day}} & 8.3 & 89 & 1.8 & -84.8 & 3.0 & -9.2 & -12.3 & -16.2 & -19.2 & -22.3 \\
 & 10.3 & 91 & 2.0 & -79.2 & 11.1 & -1.1 & -4.1 & -8.1 & -11.1 & -14.1 \\
 & 12.8 & 281 & 35.5 & -73.2 & 19.7 & 7.5 & 4.5 & 0.5 & -2.5 & -5.5 \\
 & 15.9 & 249 & 24.0 & -71.3 & 24.4 & 12.1 & 9.1 & 5.2 & 2.1 & -0.9 \\
 & 16.7 & 224 & 19.8 & -70.6 & 25.7 & 13.5 & 10.4 & 6.5 & 3.4 & 0.4 \\
\hline
\multirow{5}{*}{\textit{Night}} & 4.1 & 241 & 27.2 & -56.8 & 20.9 & 8.6 & 5.6 & 1.6 & -1.4 & -4.5 \\
 & 5.6 & 221 & 10.2 & -57.5 & 24.2 & 12.0 & 9.0 & 5.0 & 1.9 & -1.1 \\
 & 7.6 & 244 & 12.0 & -58.5 & 27.4 & 15.2 & 12.2 & 8.2 & 5.2 & 2.2 \\
 & 10.5 & 271 & 14.0 & -60.2 & 30.1 & 17.9 & 14.9 & 10.9 & 7.9 & 4.9 \\
 & 13.2 & 305 & 17.8 & -62.0 & 31.4 & 19.2 & 16.2 & 12.2 & 9.2 & 6.1 \\
\hline
\end{tabular}
\end{table*}

\section*{Acknowledgment of Sponsorship Statement}

This effort was sponsored in whole or in part by the United States Government (USG). The U.S. Government is authorized to reproduce and distribute original submissions for publication for Governmental purposes notwithstanding any copyright notation thereon.

\section*{Disclaimer}

The views and conclusions contained herein are those of the authors and should not be interpreted as necessarily representing the official policies or endorsements, either expressed or implied, of the United States Government (USG).
\bibliographystyle{IEEEtran}
\bibliography{references/references}

@book{balanisAntennaTheoryAnalysis2016,
  address = {Hoboken, NJ, USA},
  author = {Constantine A. Balanis},
  edition = {1},
  publisher = {Wiley},
  series = {New York Academy of Sciences Series},
  title = {Antenna Theory: Analysis and Design},
  year = {2016}
}

@book{carrPracticalAntennaHandbook2001,
  address = {New York, NY, USA},
  author = {Joseph J. Carr},
  edition = {4},
  publisher = {McGraw-Hill},
  title = {Practical Antenna Handbook},
  year = {2001}
}

@misc{HFRARhombic,
  author = {{AT Communication International}},
  note = {Accessed: 2026-07-05},
  title = {{AT} {HF-RA} Rhombic {HF} Antenna},
  url = {https://at-communication.com/en/hf_military_antennas_stationary/at/at_hf_ra_rhombic-antenna.html}
}

@misc{LongWireAntennasPart,
  howpublished = {\url{https://antenna2.github.io/cebik/content/wire/lw2.html}},
  title = {Long-Wire Antennas Part 2: Terminated End-Fed Long-Wire Directional Antennas}
}

@book{silverARRLAntennaBook2019,
  address = {Newington, CT, USA},
  edition = {24},
  editor = {H. Ward Silver},
  publisher = {ARRL},
  title = {The {ARRL} Antenna Book for Radio Communications},
  year = {2019}
}

@inbook{silverARRLAntennaBook2019_ch7,
  address = {Newington, CT, USA},
  title = {The {ARRL} Antenna Book for Radio Communications},
  chapter = {7},
  edition = {24},
  editor = {H. Ward Silver},
  publisher = {ARRL},
  year = {2019}
}

@inbook{silverARRLAntennaBook2019_ch11,
  address = {Newington, CT, USA},
  title = {The {ARRL} Antenna Book for Radio Communications},
  chapter = {11},
  edition = {24},
  editor = {H. Ward Silver},
  publisher = {ARRL},
  year = {2019}
}

@incollection{silverARRLAntennaBook2019_ch10,
  address = {Newington, CT, USA},
  booktitle = {The {ARRL} Antenna Book for Radio Communications},
  chapter = {10},
  edition = {24},
  editor = {H. Ward Silver},
  publisher = {ARRL},
  title = {Beverage and Traveling-Wave Antennas},
  year = {2019}
}

@book{lapinARRLHandbookRadio2024,
  address = {Newington, CT, USA},
  edition = {1},
  editor = {Gregory D. Lapin},
  publisher = {ARRL},
  title = {The {ARRL} Handbook for Radio Communications},
  year = {2024}
}

@book{stutzmanAntennaTheoryDesign2013,
  address = {Hoboken, NJ, USA},
  author = {Warren L. Stutzman and Gary A. Thiele},
  edition = {3},
  publisher = {Wiley},
  title = {Antenna Theory and Design},
  year = {2013}
}

@article{zhuMiniaturizedTransmittingLPDA2021,
  author = {Wenjun Zhu and Lixin Guo},
  doi = {10.3390/s21186034},
  note = {doi: 10.3390/s21186034},
  journal = {Sensors},
  month = sep,
  number = {18},
  pages = {6034},
  pmcid = {PMC8473130},
  pmid = {34577241},
  title = {A Miniaturized Transmitting {LPDA} Design for 2 {MHz--30} {MHz} Uses},
  volume = {21},
  year = {2021}
}

@techreport{ITU-R-F.1610-2003,
  author = {{International Telecommunication Union}},
  institution = {ITU Radiocommunication Sector (ITU-R)},
  note = {Accessed: 2025-11-15},
  number = {F.1610},
  title = {Recommendation {ITU-R} F.1610: Planning, Design and Implementation of {HF} Fixed Service Radio Systems},
  type = {Recommendation},
  url = {https://www.itu.int/rec/R-REC-F.1610/en},
  year = {2003}
}

@techreport{NTIA_Interference_Resilient_2025,
  address = {Washington, DC, USA},
  author = {{The MITRE Corporation}},
  institution = {National Telecommunications and Information Administration (NTIA)},
  month = mar,
  note = {Prepared under Contract No. 1331L523D130S0003},
  title = {Best Practices for Designing Interference-Resilient {RF} Receiving Systems},
  url = {https://www.ntia.gov/sites/default/files/2025-08/best-practices-for-designing-interference-resilient-rf-receiving-systems.pdf},
  year = {2025}
}

@techreport{ITU_R_BS_80_3_1990,
  institution = {International Telecommunication Union, Radiocommunication Sector (ITU-R)},
  address = {Geneva, Switzerland},
  type = {Recommendation},
  number = {ITU-R BS.80-3},
  note = {Originally published 1951; revisions 1978, 1986, 1990},
  title = {Transmitting Antennas in {HF} Broadcasting},
  url = {https://www.itu.int/dms_pubrec/itu-r/rec/bs/R-REC-BS.80-3-199006-I!!PDF-E.pdf},
  year = {1990}
}

@misc{DSTG_PHaRLAP,
  author = {{Defence Science and Technology Group, Commonwealth of Australia}},
  note = {Accessed: 2025-11-25},
  title = {{PHaRLAP} -- Provision of High-Frequency Raytracing Laboratory for Propagation Studies},
  url = {https://www.dst.defence.gov.au/our-technologies/pharlap-provision-high-frequency-raytracing-laboratory-propagation-studies},
  year = {2025}
}

@misc{PropLabPro_V3.2,
  author = {{Solar Terrestrial Dispatch}},
  note = {Software package for HF radio propagation and ionospheric ray-tracing},
  title = {{Proplab-Pro} {HF} Radio Propagation Laboratory ({Version} 3.2, {Build} 47 -- {August} 2025)},
  url = {https://shop.spacew.com/index.php/product/proplab-pro-hf-radio-propagation-laboratory/},
  version = {3.2 (Build 47)},
  year = {2025}
}

@misc{NTIA_LFMF,
  author = {{National Telecommunications and Information Administration (NTIA)}},
  note = {GitHub repository. Predicts basic transmission loss in the 0.01–30 MHz band over a smooth Earth with antenna heights less than 50 m.},
  title = {{LFMF}: Low Frequency/Medium Frequency Propagation Model ({C++} Implementation)},
  url = {https://github.com/NTIA/LFMF},
  version = {v1.1},
  year = {2025}
}

@misc{Sabre_XHF_2025,
  author = {{Sabre Systems, LLC}},
  howpublished = {Product brief},
  note = {Product brief; XHF-1000 (vertical) and XHF-2000 (horizontal) configurations},
  title = {Expeditionary {HF} ({XHF}) Antenna System},
  url = {https://www.sabresystems.com},
  year = {2025}
}

@article{Carr1998Beverage,
  author = {Carr, Joseph J.},
  journal = {Popular Electron.},
  month = jan,
  note = {Reprinted and archived by American Radio History},
  pages = {40--46},
  title = {The {Beverage} Antenna},
  year = {1998}
}

@online{palomarBBTDproducts,
  author = {{Palomar Engineers}},
  note = {Accessed: 2026-02-03},
  title = {Broad Band Terminated Dipoles ({BBTD}, {T2FD})},
  url = {https://palomar-engineers.com/rfi-kits/broad-band-terminated-dipoles/Antenna-Products-c21444163},
  year = {2025}
}

@inproceedings{laraway_experimental_2016,
  author = {Stephen A. Laraway and Jose Loera and Hussein Moradi and Behrouz Farhang-Boroujeny},
  booktitle = {Proc. IEEE Mil. Commun. Conf. (MILCOM)},
  doi = {10.1109/MILCOM.2016.7795412},
  note = {doi: 10.1109/MILCOM.2016.7795412},
  month = nov,
  pages = {711--716},
  title = {Experimental results of {FB-MC-SS} on a wideband {HF} {NVIS} propagation channel},
  year = {2016}
}

@article{ignatenko_wide-band_2016,
  author = {Maxim Ignatenko and Saurabh A. Sanghai and Gregor Lasser and Bradley Allen and Richard Smith and Milica Notaros and Dejan S. Filipovic},
  doi = {10.1109/MAP.2016.2609806},
  note = {doi: 10.1109/MAP.2016.2609806},
  journal = {IEEE Antennas Propag. Mag.},
  month = dec,
  number = {6},
  pages = {64--74},
  title = {Wide-band high-frequency antennas for military vehicles: Design and testing low-profile half-loop, inverted-L, and umbrella {NVIS} antennas},
  volume = {58},
  year = {2016}
}

@article{witvliet_radio_2017,
  author = {Ben A. Witvliet and Rosa Ma Alsina-Pagès},
  doi = {10.1007/s11235-017-0287-2},
  note = {doi: 10.1007/s11235-017-0287-2},
  journal = {Telecommun. Syst.},
  month = oct,
  number = {2},
  pages = {295--309},
  title = {Radio communication via near vertical incidence skywave propagation: an overview},
  volume = {66},
  year = {2017}
}

@techreport{allen_mid-latitude_nodate,
  address = {San Diego, CA},
  author = {Allen, Jeffery and Daly, Michael and Meloling, John and
Mozaffar, Nazia and Ontiveros, Marcos and Verd, Fred and Truong, Demi},
  institution = {Space and Naval Warfare Systems Center Pacific},
  number = {TR 3075 / AD1039223},
  title = {Mid-Latitude Mobile Wideband {HF-NVIS} Channel Analysis: Part 1},
  type = {Technical Report},
  url = {https://apps.dtic.mil/sti/html/tr/AD1039223/index.html},
  year = {2017}
}

@inproceedings{lamy-bergot_-air_nodate,
  address = {F{\aa}r{\"o}, Sweden},
  author = {Lamy-Bergot, Catherine and Herry, S{\'e}bastien and Bernier, Jean-Yves and Hung, Fr{\'e}d{\'e}ric Ngo Bui},
  booktitle = {Proc. 10th Nordic HF Conf. (HF 13)},
  note = {Accessed: 2025-11-26},
  title = {On-air tests results for {HF} {XL} wideband modem},
  url = {http://lamyc.free.fr/publications/NORDICHF2013_2.pdf},
  year = {2013}
}

@book{silver_arrl_2022,
  address = {Newington, CT, USA},
  editor = {H. Ward Silver},
  edition = {100},
  publisher = {ARRL},
  title = {The {ARRL} Handbook for Radio Communications 2023},
  year = {2022}
}

@incollection{silverARRLHandbook2023_ch21,
  address = {Newington, CT, USA},
  editor = {H. Ward Silver},
  booktitle = {The {ARRL} Handbook for Radio Communications 2023},
  chapter = {21},
  edition = {100},
  publisher = {ARRL},
  title = {Receiving Antennas},
  year = {2022}
}

@book{dolukhanov_propagation_1995,
  address = {Moscow, Russia},
  author = {M. Dolukhanov and Boris Kuznetsov},
  publisher = {URSS},
  title = {Propagation of Radio Waves},
  year = {1995}
}

@article{andersen_history_2017,
  author = {J. Bach Andersen},
  doi = {10.1109/MCOM.2017.7841460},
  note = {doi: 10.1109/MCOM.2017.7841460},
  journal = {IEEE Commun. Mag.},
  month = feb,
  number = {2},
  pages = {6--10},
  title = {History of communications/radio wave propagation from {M}arconi to {MIMO}},
  volume = {55},
  year = {2017}
}

@book{division_radio_2005,
  author = {{Harris Corporation, RF Communications Division}},
  lccn = {96094476},
  publisher = {Harris Corporation},
  address = {Rochester, NY, USA},
  title = {Radio Communications in the Digital Age, Volume 1: {HF} Technology},
  year = {2005}
}

@article{wang_hf_2018,
  author = {Jinlong Wang and Guoru Ding and Haichao Wang},
  doi = {10.1109/CC.2018.8456447},
  note = {doi: 10.1109/CC.2018.8456447},
  journal = {China Commun.},
  month = sep,
  number = {9},
  pages = {1--9},
  title = {{HF} Communications: Past, Present, and Future},
  volume = {15},
  year = {2018}
}

@inproceedings{adair_automatic_1989,
  address = {Boston, MA, USA},
  author = {R.T. Adair},
  booktitle = {Proc. IEEE Mil. Commun. Conf. (MILCOM)},
  doi = {10.1109/MILCOM.1989.104042},
  note = {doi: 10.1109/MILCOM.1989.104042},
  pages = {853--864},
  publisher = {IEEE},
  title = {An Automatic Link Establishment Standard for Automated Digital {HF} Radios},
  year = {1989}
}

@techreport{pinck_medium-data-rate_nodate,
  author = {L. H. Pinck and T. A. Danielson and R. North},
  title = {Medium-Data-Rate {HF} Experimental Test Results Using Modified {Harris} {RF-3201E} Transceiver and {R-2368/URR} Receiver},
  institution = {Space and Naval Warfare Systems Center},
  address = {San Diego, CA, USA},
  number = {1766},
  month = mar,
  year = {1998},
  note = {[Online]. Available: https://apps.dtic.mil/sti/pdfs/ADA349556.pdf}
}

@misc{australian_government_space_nodate,
  author = {{Australian Government Bureau of Meteorology}},
  title = {Introduction to {HF} Radio Propagation},
  note = {{Space Weather Services}},
  url = {https://www.sws.bom.gov.au/Educational/5/2/2}
}

@misc{noauthor_module_nodate,
  author = {{Naval Postgraduate School}},
  howpublished = {{NWDC} Electromagnetics Course},
  note = {Accessed: 2024-11-18},
  title = {Module 2\_3: {HF} Radiation -- Choosing the Right Frequency},
  url = {https://www.oc.nps.edu/NWDC_EM_Course/course_materials/module2_3.html}
}

@inproceedings{hoffmeyer_wideband_nodate,
  address = {Washington, DC, USA},
  author = {J. A. Hoffmeyer and M. Nesenbergs},
  booktitle = {Proc. IEEE Mil. Commun. Conf. (MILCOM)},
  month = oct,
  pages = {152--158},
  title = {Wideband {HF} modeling and simulation},
  volume = {1},
  year = {1987}
}

@article{watterson_experimental_nodate,
  author = {C. C. Watterson and J. R. Juroshek and W. D. Bensema},
  doi = {10.1109/TCOM.1970.1090438},
  note = {doi: 10.1109/TCOM.1970.1090438},
  journal = {IEEE Trans. Commun. Technol.},
  month = dec,
  number = {6},
  pages = {792--803},
  title = {Experimental confirmation of an {HF} channel model},
  volume = {18},
  year = {1970}
}

@inproceedings{dhar_equalized_1982,
  author = {S. Dhar and B. D. Perry},
  booktitle = {Proc. IEEE Mil. Commun. Conf. (MILCOM)},
  doi = {10.1109/MILCOM.1982.4805973},
  note = {doi: 10.1109/MILCOM.1982.4805973},
  month = oct,
  pages = {29.5--1--29.5--5},
  title = {Equalized Megahertz-Bandwidth {HF} Channels for Spread Spectrum Communications},
  year = {1982}
}

@article{skaug_experiment_1984,
  author = {R. Skaug},
  doi = {10.1049/ip-f-1.1984.0015},
  note = {doi: 10.1049/ip-f-1.1984.0015},
  journal = {IEE Proc. F Commun. Radar Signal Process.},
  number = {1},
  pages = {87},
  title = {Experiment with Spread Spectrum Modulation on an {HF} Channel},
  volume = {131},
  year = {1984}
}

@inproceedings{lossmann_hf_2011,
  address = {Torino, Italy},
  author = {E. Lossmann and M.-A. Meister and U. Madar},
  booktitle = {Proc. IEEE-APS Top. Conf. Antennas Propag. Wireless Commun. (APWC)},
  doi = {10.1109/APWC.2011.6046802},
  note = {doi: 10.1109/APWC.2011.6046802},
  month = sep,
  pages = {812--814},
  publisher = {IEEE},
  title = {On {HF} Communication Link Parameter Estimation in the Baltic Region},
  year = {2011}
}

@book{itziar_angulo_handbook_nodate,
  address = {Geneva, Switzerland},
  author = {I. Angulo and L. Barclay and Y. Chernov and N. DeMinco and I. Fern{\'{a}}ndez and U. Gil and D. Guerra and J. Milsom and I. Pe{\~n}a and D. {De La Vega}},
  publisher = {Int. Telecommun. Union (ITU)},
  title = {Handbook on Ground Wave Propagation},
  year = {2014}
}

@book{fabrizio_high_2013,
  address = {New York, NY, USA},
  author = {Giuseppe Aureliano Fabrizio},
  edition = {1},
  publisher = {McGraw-Hill Education},
  title = {High Frequency Over-the-Horizon Radar: Fundamental Principles, Signal Processing, and Practical Applications},
  year = {2013}
}

@book{balanis_balanis_2024,
  address = {Hoboken, NJ, USA},
  author = {Constantine A. Balanis},
  doi = {10.1002/9781394180042},
  note = {doi: 10.1002/9781394180042},
  edition = {3},
  publisher = {Wiley},
  title = {Balanis' Advanced Engineering Electromagnetics},
  year = {2024}
}

@book{jin_theory_2010,
  address = {Hoboken, NJ, USA},
  author = {Jian-Ming Jin},
  publisher = {Wiley},
  title = {Theory and Computation of Electromagnetic Fields},
  year = {2010}
}

@article{leferink_man-made_2012,
  author = {Frank Leferink and François Silva and Johan Catrysse and Sven Battermann and Véronique Beauvois and Anne Roc'h},
  journal = {Radio Sci. Bull.},
  month = jan,
  title = {Man-made noise in our living environments},
  volume = {334},
  year = {2012}
}

@inproceedings{sorecau_man-made_2022,
  address = {Timisoara, Romania},
  author = {Emil Sorecau and Mirela Sorecau and Neculai Craiu and Annamaria Sarbu and Paul Bechet},
  booktitle = {Proc. Int. Symp. Electron. Telecommun. (ISETC)},
  doi = {10.1109/ISETC56213.2022.10010035},
  note = {doi: 10.1109/ISETC56213.2022.10010035},
  month = nov,
  pages = {1--4},
  publisher = {IEEE},
  title = {Man-made Noise Measurement System for {HF} Band Based on {SDR} Platforms - Design and Implementation},
  year = {2022}
}

@article{noauthor_recommendation_nodate-1,
  journal = {International Telecommunication Union (ITU) Recommendation},
  month = aug,
  note = {Accessed: 2024-12-01},
  title = {Recommendation {ITU-R} P.372-17 (08/2024) -- Radio Noise},
  url = {https://www.itu.int/dms_pubrec/itu-r/rec/p/R-REC-P.372-17-202408-I!!PDF-E.pdf},
  year = {2024}
}

@misc{noauthor_background_2017,
  month = dec,
  note = {Radio Society of Great Britain (RSGB) leaflet},
  title = {The Background Noise on the {HF} Amateur Bands},
  url = {https://rsgb.org/main/files/2017/12/221216-Noise-leaflet-issue-2.pdf},
  year = {2017}
}

@techreport{noauthor_interoperability_2011,
  address = {Washington, DC, USA},
  author = {{U.S. Department of Defense}},
  institution = {U.S. Dept. of Defense},
  month = dec,
  note = {Accessed: Nov. 26, 2025},
  number = {MIL-STD-188-141C},
  title = {Interoperability and Performance Standards for Medium and High Frequency Radio Systems},
  type = {Mil. Std.},
  url = {https://hflink.com/standards/MIL_STD_188-141C.pdf},
  year = {2011}
}

@article{rady_how_2024,
  author = {Mina Rady and Oana Iova and Hervé Rivano and Angeliki Deligianni and Leonidas Drikos},
  doi = {10.1016/j.adhoc.2024.103418},
  note = {doi: 10.1016/j.adhoc.2024.103418},
  journal = {Ad Hoc Netw.},
  month = apr,
  pages = {103418},
  title = {How does {Wi-Fi} 6 fare? An industrial outdoor robotic scenario},
  volume = {156},
  year = {2024}
}

@incollection{sevgi_ground_2003,
  address = {Piscataway, NJ, USA},
  author = {Levent Sevgi and Funda Akleman},
  booktitle = {Complex Electromagnetic Problems and Numerical Simulation Approaches},
  pages = {63--125},
  publisher = {IEEE Press},
  title = {Ground Wave Propagation},
  year = {2003}
}

@article{callaway_gray_nodate,
  author = {E. Callaway},
  journal = {QEX},
  month = {Nov./Dec.},
  note = {Accessed: 2026-07-05},
  title = {Gray Line Propagation, or {Florida} to {Cocos} ({Keeling}) on 80 m},
  url = {https://www.arrl.org/files/file/QEX%20Binaries/2016/Callaway.pdf},
  year = {2016}
}

@article{raouafi_parker_2023,
  author = {N. E. Raouafi and others},
  doi = {10.1007/s11214-023-00952-4},
  note = {doi: 10.1007/s11214-023-00952-4},
  journal = {Space Sci. Rev.},
  number = {1},
  pages = {8},
  title = {Parker Solar Probe: Four Years of Discoveries at Solar Cycle Minimum},
  volume = {219},
  year = {2023}
}

@article{domingo_soho_1995,
  author = {V. Domingo and B. Fleck and A. I. Poland},
  doi = {10.1007/BF00768758},
  note = {doi: 10.1007/BF00768758},
  journal = {Space Sci. Rev.},
  number = {1},
  pages = {81--84},
  title = {{SOHO}: The Solar and Heliospheric Observatory},
  volume = {72},
  year = {1995}
}

@inproceedings{kaiser_stereo_2007,
  address = {Big Sky, MT, USA},
  author = {Michael L. Kaiser and W. James Adams},
  booktitle = {Proc. IEEE Aerosp. Conf.},
  doi = {10.1109/AERO.2007.352745},
  note = {doi: 10.1109/AERO.2007.352745},
  pages = {1--8},
  publisher = {IEEE},
  title = {Stereo Mission Overview},
  year = {2007}
}

@article{priest_magnetic_2002,
  author = {E. R. Priest and T. G. Forbes},
  doi = {10.1007/s001590100013},
  note = {doi: 10.1007/s001590100013},
  journal = {Astron. Astrophys. Rev.},
  number = {4},
  pages = {313--377},
  title = {The Magnetic Nature of Solar Flares},
  volume = {10},
  year = {2002}
}

@article{solanki_sunspots_2003,
  author = {Sami K. Solanki},
  doi = {10.1007/s00159-003-0018-4},
  note = {doi: 10.1007/s00159-003-0018-4},
  journal = {Astron. Astrophys. Rev.},
  number = {2},
  pages = {153--286},
  title = {Sunspots: An Overview},
  volume = {11},
  year = {2003}
}

@article{lakhina_geomagnetic_2016,
  author = {Gurbax S. Lakhina and Bruce T. Tsurutani},
  doi = {10.1186/s40562-016-0037-4},
  note = {doi: 10.1186/s40562-016-0037-4},
  journal = {Geosci. Lett.},
  number = {1},
  pages = {5},
  title = {Geomagnetic Storms: Historical Perspective to Modern View},
  volume = {3},
  year = {2016}
}

@article{danilov2001effects,
  author = {A. D. Danilov and J. Lastovicka},
  journal = {Int. J. Geomagn. Aeron.},
  number = {3},
  pages = {209--224},
  publisher = {American Geophysical Union},
  title = {Effects of Geomagnetic Storms on the Ionosphere and Atmosphere},
  volume = {2},
  year = {2001}
}

@online{FCC97_3_a_8,
  title = {Definitions, 47 {C.F.R.} {\S}~97.3},
  url = {https://www.ecfr.gov/current/title-47/part-97/section-97.3}
}

@techreport{institute_for_telecommunication_sciences_required_1969,
  address = {Boulder, CO, USA},
  author = {Wesley M. Beery and Gene G. Ax and H. Akima},
  institution = {Institute for Telecommunication Sciences, ESSA Research Laboratories},
  type = {ESSA Technical Report},
  number = {ERL 131-ITS 92},
  title = {Required Signal-to-Noise Ratios for {HF} Communication Systems},
  url = {https://catalog.hathitrust.org/Record/102381899},
  year = {1969}
}

@online{noauthor_amateur_2023,
  author = {{Federal Communications Commission}},
  howpublished = {Federal Register, vol. 88, no. 234, pp. 85126--85129},
  month = dec,
  year = {2023},
  note = {Accessed: 2026-07-05},
  title = {Amateur Radio Service Rules To Permit Greater Flexibility in Data Communications},
  url = {https://www.govinfo.gov/content/pkg/FR-2023-12-07/pdf/2023-26770.pdf}
}

@online{noauthor_image_nodate,
  author = {M. Bruchanov},
  title = {Image Communication on Short Waves -- {SSTV}, {WEFAX}, {HamDRM}},
  url = {https://www.sstv-handbook.com/}
}

@online{noauthor_pactor-4_nodate,
  author = {{SCS Special Communications Systems GmbH}},
  title = {{PACTOR-4}},
  url = {https://www.p4dragon.com/pactor-4.html}
}

@article{klapashchuk_analysis_2024,
  author = {Ihor Klapashchuk and Andrii Veryha},
  doi = {10.31861/sisiot2024.2.02005},
  note = {doi: 10.31861/sisiot2024.2.02005},
  journal = {Secur. Infocommun. Syst. Internet Things},
  number = {2},
  pages = {02005},
  title = {Analysis of Amateur Radio Frequency Code Modulation Protocols for Transmitting Short Messages},
  volume = {2},
  year = {2024}
}

@online{noauthor_psk31_nodate,
  author = {{American Radio Relay League}},
  title = {{PSK31} Specification},
  url = {https://www.arrl.org/psk31-spec}
}

@online{noauthor_mfsk_nodate,
  author = {{American Radio Relay League}},
  title = {{MFSK} Specification},
  url = {https://www.arrl.org/mfsk-spec}
}

@techreport{wagner_wideband_nodate,
  author      = {Leonard S. Wagner and Joseph A. Goldstein and Eather A. Chapman},
  title       = {Wideband {HF} Channel Prober: System Description},
  institution = {Naval Research Laboratory},
  number      = {NRL Report 8622, AD-A127-040},
  address     = {Washington, DC, USA},
  year        = {1983}
}

@online{sun_rotation,
  month = sep,
  howpublished = {{HMI} Science Nuggets, Stanford University},
  title = {How to Keep the Sun's Equator Rotating Faster than its Poles: Giant Cells},
  url = {http://hmi.stanford.edu/hminuggets/?p=715},
  year = {2019}
}

@online{sunspot_numbers,
  author = {{NOAA National Geophysical Data Center}},
  title = {Sunspot Numbers},
  url = {https://www.ngdc.noaa.gov/stp/iono/sunspot.html}
}

@online{f10p7_RadioEmmissions,
  author = {{NOAA Space Weather Prediction Center}},
  title = {{F10.7} cm Radio Emissions},
  url = {https://www.swpc.noaa.gov/phenomena/f107-cm-radio-emissions}
}

@article{wang_MUFcomparison_2024,
  author = {Wang, Jian and Han, Han and Shi, Yafei and Yang, Cheng and Liu, Yiran and Wang, Zequan},
  doi = {10.1016/j.asr.2024.05.060},
  note = {doi: 10.1016/j.asr.2024.05.060},
  journal = {Adv. Space Res.},
  month = sep,
  number = {5},
  pages = {2452--2462},
  title = {Comparison and validation of {MOF} observations and {MUF} predictions from seven different models},
  volume = {74},
  year = {2024}
}

@online{NOAA_SolarCycleProgression,
  author = {{NOAA Space Weather Prediction Center}},
  title = {Solar Cycle Progression},
  url = {https://www.swpc.noaa.gov/products/solar-cycle-progression}
}

@techreport{hakura1968polar,
  address = {Greenbelt, MD, USA},
  author = {Hakura, Yukio},
  institution = {National Aeronautics and Space Administration},
  month = jun,
  number = {NASA-TN-D-4473},
  title = {Polar cap absorptions and associated solar terrestrial events throughout the 19th solar cycle},
  type = {NASA Technical Note},
  url = {https://ntrs.nasa.gov/citations/19680016408},
  year = {1968}
}

@article{uryadov_impact_2018,
  author = {Uryadov, V.P. and Vybornov, F.I. and Kolchev, A.A. and Vertogradov, G.G. and Sklyarevsky, M.S. and Egoshin, I.A. and Shumaev, V.V. and Chernov, A.G.},
  doi = {10.1016/j.asr.2017.07.003},
  note = {doi: 10.1016/j.asr.2017.07.003},
  journal = {Adv. Space Res.},
  month = apr,
  number = {7},
  pages = {1837--1849},
  title = {Impact of heliogeophysical disturbances on ionospheric {HF} channels},
  volume = {61},
  year = {2018}
}

@article{hargreaves_new_2005,
  author = {Hargreaves, J. K.},
  doi = {10.5194/angeo-23-359-2005},
  note = {doi: 10.5194/angeo-23-359-2005},
  journal = {Ann. Geophys.},
  month = feb,
  number = {2},
  pages = {359--369},
  title = {A new method of studying the relation between ionization rates and radio-wave absorption in polar-cap absorption events},
  volume = {23},
  year = {2005}
}

@article{kotaki_global_1984,
  author = {Kotaki, Minoru},
  doi = {10.1016/0021-9169(84)90026-6},
  note = {doi: 10.1016/0021-9169(84)90026-6},
  journal = {J. Atmos. Terr. Phys.},
  month = oct,
  number = {10},
  pages = {867--877},
  title = {Global distribution of atmospheric radio noise derived from thunderstorm activity},
  volume = {46},
  year = {1984}
}

@online{australia_spaceWeather,
  author = {{Australian Government Bureau of Meteorology}},
  title = {Space Weather Services Website},
  url = {https://www.sws.bom.gov.au/Educational/1/2/5}
}

@misc{madrigal_db,
  author = {W. Rideout and K. Cariglia},
  note = {Accessed: 2025-11-25},
  title = {{CEDAR} {Madrigal} Database},
  url = {https://cedar.openmadrigal.org},
  year = {2024}
}

@article{cervera2014,
  author = {M. A. Cervera and T. J. Harris},
  doi = {10.1002/2013JA019247},
  note = {doi: 10.1002/2013JA019247},
  journal = {J. Geophys. Res. Space Phys.},
  number = {1},
  pages = {431--440},
  title = {Modeling ionospheric disturbance features in quasi-vertically incident ionograms using {3-D} magnetoionic ray tracing and atmospheric gravity waves},
  volume = {119},
  year = {2014}
}

@article{bilitza2022,
  author = {D. Bilitza and M. Pezzopane and V. Truhlik and D. Altadill and B. W. Reinisch and A. Pignalberi},
  doi = {10.1029/2022RG000792},
  note = {doi: 10.1029/2022RG000792},
  journal = {Rev. Geophys.},
  number = {4},
  pages = {e2022RG000792},
  title = {The {International} {Reference} {Ionosphere} model: A review and description of an ionospheric benchmark},
  volume = {60},
  year = {2022}
}

@article{schneider2018shortwave,
  author = {D. Schneider},
  doi = {10.1109/MSPEC.2018.8389174},
  note = {doi: 10.1109/MSPEC.2018.8389174},
  journal = {IEEE Spectr.},
  month = jul,
  number = {7},
  pages = {12},
  title = {Wall {Street} tries shortwave radio},
  volume = {55},
  year = {2018}
}

@inproceedings{ma2025overair,
  author = {R. Ma and S. Wang and H. Topozlu and E. Berger and D. Ludois and N. Behdad},
  title = {Over-the-Air Transmission of Wideband, High-Order Digitally Modulated Waveforms at the {HF} Band Using a Non-{LTI} Electrically Small Antenna with Enhanced Bandwidth-Efficiency Product},
  booktitle = {Proc. IEEE CNC-USNC-URSI North Amer. Radio Sci. Meeting (Joint AP-S Symp.)},
  address = {Ottawa, ON, Canada},
  pages = {3309--3309},
  year = {2025},
  doi = {10.23919/CNC-USNC-URSI64444.2025.11419990},
  note = {doi: 10.23919/CNC-USNC-URSI64444.2025.11419990}
}

@misc{ludois2026qmod,
  author = {D. Ludois and N. Behdad and M. Liben and M. Mirmozafari},
  title = {Antenna System with Quality-Factor Modulation},
  howpublished = {U.S. Patent 12{,}646{,}847 B2},
  note = {Issued Jun. 2, 2026},
  year = {2026}
}

@article{franke2020,
  author = {S. J. Franke and B. Somerville and J. Taylor},
  journal = {QEX},
  month = {Jul./Aug.},
  note = {[Online]. Available: https://wsjt.sourceforge.io/FT4\_FT8\_QEX.pdf},
  pages = {39--45},
  title = {The {FT4} and {FT8} Communication Protocols},
  year = {2020}
}

@article{reinisch2011giro,
  author = {B. W. Reinisch and I. A. Galkin},
  doi = {10.5047/eps.2011.03.001},
  note = {doi: 10.5047/eps.2011.03.001},
  journal = {Earth Planets Space},
  pages = {377--381},
  title = {Global Ionospheric Radio Observatory ({GIRO})},
  volume = {63},
  year = {2011}
}

@article{whitehead1989,
  author = {J. D. Whitehead},
  doi = {10.1016/0021-9169(89)90122-0},
  note = {doi: 10.1016/0021-9169(89)90122-0},
  journal = {J. Atmos. Terr. Phys.},
  number = {5},
  pages = {401--424},
  title = {Recent work on mid-latitude and equatorial sporadic-{E}},
  volume = {51},
  year = {1989}
}

\begin{IEEEbiography}[{\includegraphics[width=1in,height=1.25in,clip,keepaspectratio]{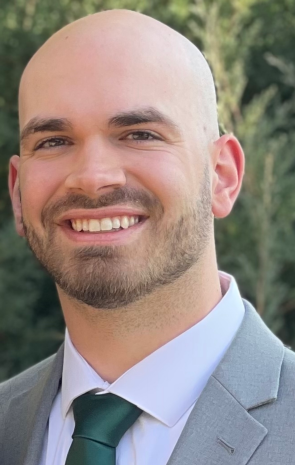}}]{Eric Weber}{\,} received the Ph.D. degree in electrical engineering from the University of Wisconsin Madison, Madison, WI, USA, in 2026. His research interests include microwave and radio frequency engineering, microwave electronic devices, medical imaging, and bioelectromagnetics.
\end{IEEEbiography}

\begin{IEEEbiography}[{\includegraphics[width=1in,height=1.25in,clip,keepaspectratio]{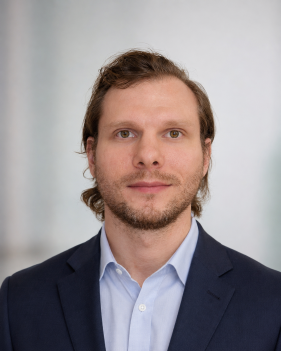}}]{Ted Nowak}{\,} received his bachelor's degree from Marquette University and is currently pursuing his Ph.D. in electrical engineering at the University of Wisconsin–Madison. His research interests include RF, microwave, and millimeter-wave engineering, biomedical electromagnetics, antenna and resonator design, medical imaging, and RF instrumentation. His doctoral research focuses on the development of instrumentation and electromagnetic systems for electron paramagnetic resonance (EPR) imaging.
\end{IEEEbiography}

\begin{IEEEbiography}[{\includegraphics[width=1in,height=1.25in,clip,keepaspectratio]{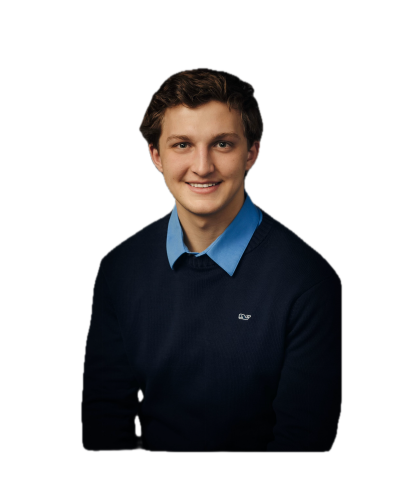}}]{Joseph Berg}{\,} earned the B.S. degrees in electrical engineering and computer science (2024) and M.S. degree in electrical engineering (2025) from the University of Wisconsin – Madison. He is currently pursuing the Ph.D. degree in Electrical Engineering at UW-Madison, where his research focuses on non-LTI antennas, phased array antennas, and HF propagation.
\end{IEEEbiography}

\begin{IEEEbiography}[{\includegraphics[width=1in,height=1.25in,clip,keepaspectratio]{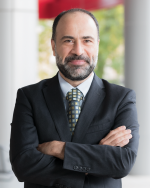}}]{Nader Behdad}{\,} (S’98-M’06-SM’12-F’17) received the B.S. degree in Electrical Engineering from Sharif University of Technology in 2000 and the M.S. and Ph.D. degrees in Electrical Engineering from University of Michigan-Ann Arbor in 2003 and 2006, respectively. Currently he holds the Harvey D. Spangler and the Vilas Distinguished Achievement Professorships in the Department of Electrical and Computer Engineering of the University of Wisconsin-Madison. His research expertise is in the area of applied electromagnetics with particular focus on electrically small antennas, phased-array antennas, microwave periodic structures, high-power microwaves, and biomedical applications of RF and microwaves. He has 25 issued U.S. patents in these areas, with three additional patent applications filed with the USPTO. Dr. Behdad has served as a consultant on topics related to designing antennas and phased arrays for industry. He has also served as a consultant and an expert witness for different U.S. law firms on topics related to intellectual property disputes as well as cell phone record analysis and historical cell site analysis. Over the years, his research has been sponsored by various U.S. Federal agencies including the U.S. Navy, U.S. Air Force, U.S. Army, National Science Foundation, and the Defense Health Agency among others.

Dr. Behdad has graduated 33 Ph.D. and 16 M.S. students so far and served as the research advisor of 32 other post-doctoral research fellows and visiting scholars. He is the recipient of the 2025 John Kraus Antenna Award, the 2021 H. A. Wheeler Prize Paper Award, the 2014 R. W. P. King Prize Paper Award, and the 2012 Piergiorgio L. E. Uslenghi Letters Prize Paper Award of the IEEE Antennas and Propagation Society. He also received the Byron Bird Award for Excellence in a Research Publication, McFarland-Bascom Professorship, Harvey D. Spangler Faculty Scholar Award, the H. I. Romnes Faculty Award, and the Vilas Associates Award from the University of Wisconsin-Madison. In 2011, Dr. Behdad received the CAREER award from the U.S. National Science Foundation, the Young Investigator Award from the United States Air Force Office of Scientific Research, and the Young Investigator Award from the United States Office of Naval Research. He served as a member of the Fellow Election Committee of IEEE Nuclear and Plasma Sciences Society (2022-2025) and served as the 2020 chair of the paper awards committee of the IEEE Antennas and Propagation Society. He also served as an Associate Editor for IEEE Antennas and Wireless Propagation Letters (2011-2015) and as the co-chair of the technical program committee of the 2012 IEEE International Symposium on Antennas and Propagation and USNC/URSI National Radio Science Meeting.
\end{IEEEbiography}

\end{document}